\documentclass[aps,superscriptaddress,floatfix,twocolumn,footinbib,prb,longbibliography]{revtex4-2}
\usepackage{amssymb}
\usepackage{amsmath}
\usepackage{amsfonts}
\usepackage{appendix}
\usepackage{bm}
\usepackage{graphicx}
\usepackage{epsfig}
\usepackage{epstopdf}
\usepackage{balance}
\usepackage[dvipsnames]{xcolor}
\usepackage{calc}
\usepackage{natbib}
\usepackage[colorlinks,
            linkcolor=blue,
            anchorcolor=blue,
            citecolor=blue,
            urlcolor=blue]{hyperref}
\usepackage{lipsum}
\usepackage[version=3]{mhchem} 

\usepackage{color}
\usepackage{soul}
\usepackage[etex=true,export]{adjustbox}
\usepackage{makecell}
\usepackage{multirow}
\usepackage{tabularx,multirow,array,diagbox}
\usepackage{adjustbox}
\usepackage{booktabs}
\begin{document}
\title{Interaction-driven topological phase diagram of twisted bilayer MoTe$_2$ }

\author{Wen-Xuan Qiu}
\affiliation{School of Physics and Technology, Wuhan University, Wuhan 430072, China}
\author{Bohao Li}
\affiliation{School of Physics and Technology, Wuhan University, Wuhan 430072, China}
\author{Xun-Jiang Luo}
\affiliation{School of Physics and Technology, Wuhan University, Wuhan 430072, China}
\author{Fengcheng Wu}
\email{wufcheng@whu.edu.cn}
\affiliation{School of Physics and Technology, Wuhan University, Wuhan 430072, China}
\affiliation{Wuhan Institute of Quantum Technology, Wuhan 430206, China}

\begin{abstract}
Twisted bilayer MoTe$_2$ is a promising platform to investigate the interplay between band topology and many-body interaction. We present a theoretical study of its interaction-driven quantum phase diagrams based on a three-orbital model, which can be viewed as a generalization of the Kane-Mele-Hubbard model with one additional orbital and long-range Coulomb repulsion. We predict a cascade of phase transitions tuned by the twist angle $\theta$. At the hole filling factor $\nu=1$ (one hole per moir\'e unit cell), the ground state can be in the  multiferroic phase with coexisting spontaneous layer polarization and magnetism, the quantum anomalous Hall phase, and finally the topologically trivial magnetic phases, as $\theta$ increases from $1.5^{\circ}$ to $5^{\circ}$. At $\nu=2$, the ground state can have a second-order phase transition between an antiferromagnetic phase and the quantum spin Hall phase as $\theta$ passes through a critical value. The dependence of the phase boundaries on model parameters such as the gate-to-sample distance, the dielectric constant, and the moir\'e potential amplitude is examined. The predicted phase diagrams can guide the search for topological phases in twisted transition metal dichalcogenide homobilayers. 
\end{abstract}

\maketitle
\section{introduction}
Moir\'e superlattices formed by group-VI transition metal dichalcogenides (TMD) provide a new solid-state platform to simulate model Hamiltonians on honeycomb and triangular lattices \cite{Wu2018,Wu2019,angeli2021gamma,zhangyang_electronic,Xian_Realization2021, Naik2018}.  TMD monolayers are semiconductors with sizable spin-orbit splitting in the valence band \cite{xiao2012}, and therefore, have a reduced number of low-energy degrees of freedom compared to monolayer graphene. This theoretical simplification makes moir\'e TMD  systems an appealing platform for quantum simulation.  In twisted TMD homobilayers, the low-energy moir\'e valence bands can be mapped to the Kane-Mele model on a honeycomb lattice for $\pm K$ valley states \cite{Wu2019}, or a honeycomb tight-binding model with negligible spin-orbit coupling for $\Gamma$ valley states \cite{angeli2021gamma,zhangyang_electronic,Xian_Realization2021}. A variety of physical effects have been theoretically predicted for the twisted TMD homobilayers \cite{Wu2019,angeli2021gamma,zhangyang_electronic,Xian_Realization2021,Hongyi2019_giant,Pan2020,Pan2020a, Devakul2021,Heqiu_spontaneous2021,  Kaushal2022_magnetic,Mikael2022_Topological, Valentin_anomalous2022,Abouelkomsan_multiferroicity2022, Kaushal2023_Kanamori}. Experimental studies have reported correlated insulating states as well as signatures of interaction-induced topological states in twisted bilayer WSe$_2$ ($t$WSe$_2$) \cite{Wang2020, Ghiotto2021_Quantum, Foutty_mapping2023} and twisted bilayer MoTe$_2$ ($t$MoTe$_2$) \cite{Xiaodong2023a,Xiaodong2023b,Yihang2023_integer}. In moir\'e TMD heterobilayers, low-energy states reside primarily in one layer due to the band offset between two different TMD materials and can simulate the generalized Hubbard model on triangular lattices \cite{Wu2018}. Experimental observations of Mott insulators \cite{Tang2020, Regan2020},  generalized Wigner crystals \cite{Xu2020,Li2021a}, metal-insulator transitions \cite{Li2021_Continuous, Mingjie_Tuning2022}, and heavy fermion behaviors \cite{Zhao2023_Gate} in moir\'e TMD heterobilayers support the Hubbard model physics.  An out-of-plane electric field can reduce the band offset in the heterbilayer and induce band inversion, which can result in topological moir\'e bands \cite{Zhang2021}. The electric field-induced topological states have been experimentally realized in AB-stacked MoTe$_2$/WSe$_2$ heterobilayer \cite{Li2021,Zhao2022,Tao2022}, which stimulated active theoretical studies on the nature of the states \cite{Pan2022,Xie2022a, Devakul2022,Chang2022,Dong2022,Rademaker2022,Daniele_Chiral2023,Xie2022,Xie2022c,Xunjiang2022}. 
\begin{figure}[t]
\centering
\includegraphics[width=1\columnwidth]{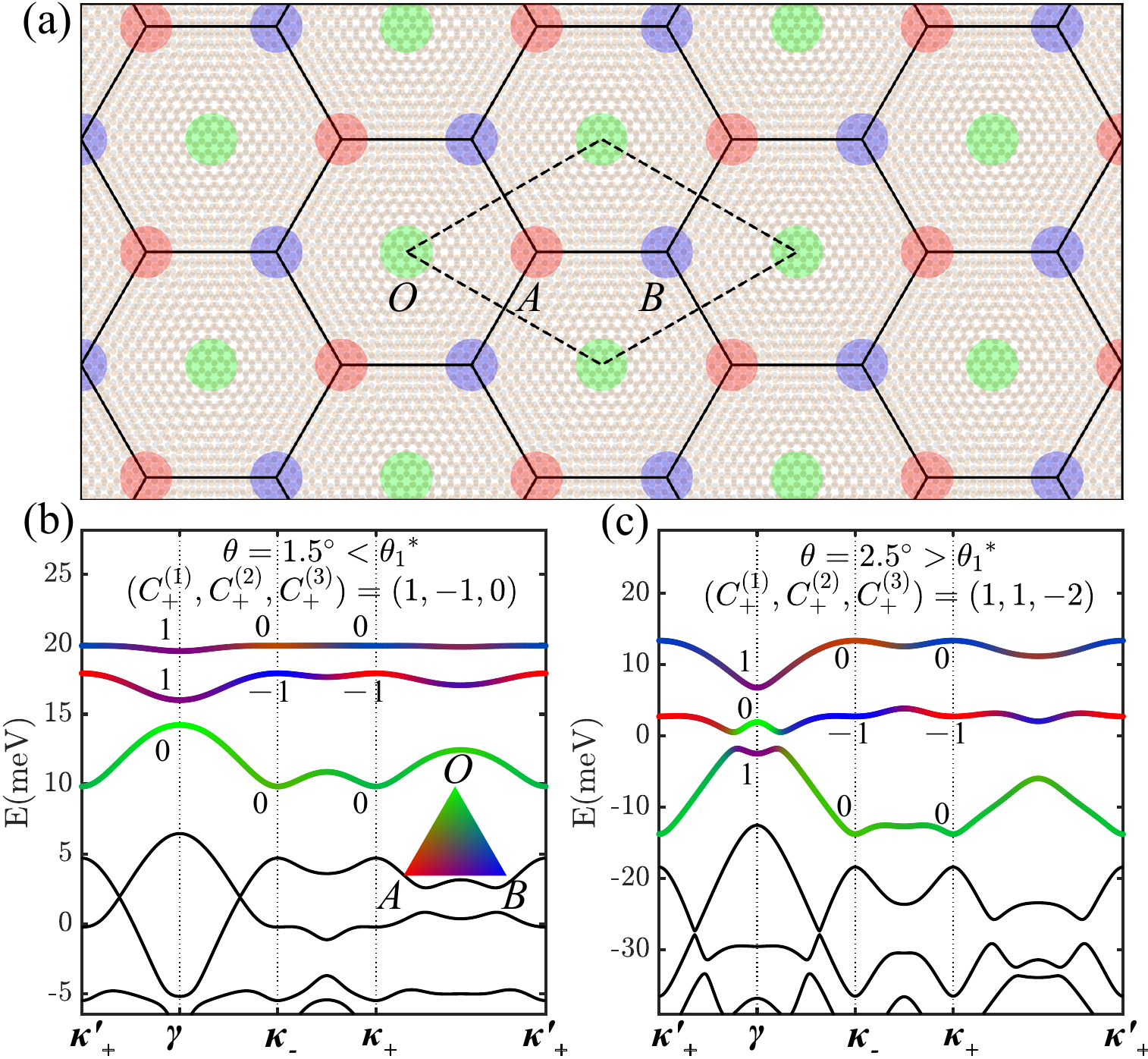}
\caption{(a) Moir\'e superlattices formed in $t$MoTe$_2$. The dots with green, red, and blue colors indicate, respectively, $O$, $A$, and $B$ sites, which also correspond to the centers of the three Wannier orbitals. The black dashed lines mark a moir\'e unit cell. (b), (c) Moir\'e band structure at $\theta=1.5^{\circ}$ and $2.5^{\circ}$ in $+K$ valley. The integer numbers at $\bm{\gamma}$ and $\bm{\kappa}_{\pm}$ label the $C_{3z}$ angular momentum of the first three bands (see Appendix \ref{appa}), while the color represents the relative weight of the three Wannier orbitals. The band structures are plotted along high-symmetry paths in momentum space with $\bm{\kappa}'_+=-\bm{\kappa}_-$. }
\label{fig1}
\end{figure}

In this paper, we present a theoretical study of the interaction-driven quantum phase diagram of $t$MoTe$_2$, where the two layers are rotated by an angle $\theta$ relative to the rhombohedral-stacked bilayer. As predicted in Ref.~\onlinecite{Wu2019}, low-energy moir\'e valence bands of $t$MoTe$_2$ originate from $\pm K$ valleys of monolayer MoTe$_2$, where $\pm K$ refer to the monolayer Brillouin zone corners. The intralayer moir\'e potentials confine low-energy states to two different locations [i.e., $A$ and $B$ sites of the moir\'e superlattice shown in Fig.~\ref{fig1}(a)] in the bottom and top layers, where the two locations exactly correspond to the two sublattices of a honeycomb lattice. The rotation between the two layers generates layer-dependent momentum shifts in the kinetic energy, which induces effective valley-contrasting and sublattice-dependent fluxes on the honeycomb lattice model. Because of the lattice geometry and the flux pattern, the first two moir\'e valence bands realize the Haldane model \cite{Haldane1988} within one valley and the Kane-Mele model \cite{Kane2005,Kane2005a} when both valleys are considered \cite{Wu2019}. Moreover, the moir\'e bands have narrow bandwidth, and electron interaction effects are enhanced. Therefore,  $t$MoTe$_2$ can serve as a model system to study the interplay between band topology and many-body interaction. However, $t$MoTe$_2$ has not been studied experimentally until very recently \cite{Xiaodong2023a,Xiaodong2023b,Yihang2023_integer}. Two independent experiments reported  signatures of integer and fractional quantum anomalous Hall states in $t$MoTe$_2$ with $\theta \approx 3.7^{\circ}$ \cite{Xiaodong2023b} and $3.4^{\circ}$ \cite{Yihang2023_integer}, respectively. This exciting experimental progress makes the theoretical prediction of Ref.~\onlinecite{Wu2019} to be of experimental relevance, and calls for thorough studies of the many-body interaction effects in $t$MoTe$_2$.

The mapping between the first two moir\'e bands and the Kane-Mele model allows to study the interaction effects within a lattice model, which should be a generalized Kane-Mele-Hubbard model since the realistic interaction is the long-range Coulomb repulsion. However, such a mapping is limited to the small $\theta$ regime of $\theta<\theta_1^*$, where $\theta_1^*$ is estimated to be about $1.74^{\circ}$ in Ref.~\onlinecite{Wu2019}. For $\theta>\theta_1^*$, there is a topological band inversion between the second and the third bands. To describe the low-energy moir\'e bands in a unified manner regardless of the exact values of $\theta$, we construct a three-orbital model based on the Wannier states of the first three bands. The three-orbital model faithfully captures the energy, symmetry, and topology of the first three bands. The interacting model is then obtained by projecting the Coulomb interaction onto the Bloch-like states derived from the Wannier orbitals. The interacting Hamiltonian is formulated in momentum space and implicitly includes all microscopic interactions such as density-density interaction and Hund's coupling.

We calculate the quantum phase diagrams of the interacting three-orbital model at hole filling factors $\nu=1$ and $\nu=2$, where $\nu$ counts the number of doped holes per moir\'e unit cell. The phase diagrams are obtained by comparing the energy of multiple competing states within self-consistent Hartree-Fock approximation. We construct the phase diagram as a function of $\theta$ and examine its dependence on the gate-to-sample distance $d$, the dielectric constant $\epsilon$, and the moir\'e potential amplitude $V$. We predict a cascade of phase transitions tuned by $\theta$ in the range between $1.5^\circ$ and $5^\circ$. At $\nu=1$, the ground state is in the multiferroic phase with spontaneous layer polarization and magnetism for small $\theta$, the quantum anomalous Hall phase with a Chern number of 1 for intermediate $\theta$, and topologically trivial magnetic phases for large $\theta$. At $\nu=2$, the ground state is in interaction-driven magnetic phases for small $\theta$ and turns into the quantum spin Hall phase for large $\theta$. We also discuss how the phase boundaries depend on parameters such as $d$, $\epsilon$, and $V$.  
The predicted phase diagrams demonstrate the remarkable richness of $t$MoTe$_2$, and are expected to guide further experimental and theoretical studies.

There are several benefits of doing the many-body calculation using the three-orbital model compared to the full continuum model.
First, the constructed Wannier states provide a clear real-space picture of the low-energy degrees of freedom and valuable intuition for studying interaction physics. The localized Wannier orbitals allow us to devise various mean-field states with different layer polarization, magnetization, and topology. Second, the symmetry and topology of electronic states become more apparent in the Wannier basis. Third, the three-orbital model can be further studied using techniques (e.g., exact diagonalization) beyond mean-field theory.

The paper is organized as follows. In Sec.~\ref{sec1}, we construct the noninteracting as well as the interacting three-orbital model.  Construction of the Wannier states is informed by symmetry analysis of the moir\'e bands, which is detailed in Appendix~\ref{appa}. 
In Sec.~\ref{sec2}, we present the phase diagrams at $\nu=1$ and $\nu=2$. Additional discussion on the phase diagram and the model is given, respectively, in Appendices~\ref{appb} and \ref{appc}.
In Sec.~\ref{sec3}, we conclude with a summary and a discussion. Our theoretical results are compared with the recent experimental findings, and future directions are discussed.

\section{Model}\label{sec1}
\subsection{Moir\'e Hamiltonian}
The moir\'e superlattices of $t$MoTe$_2$ have $D_{3}$ point-group symmetry generated by a threefold rotation $C_{3z}$ around the out-of-plane $\hat{z}$ axis and a twofold rotation $C_{2y}$ around the in-plane $\hat{y}$ axis that exchanges the bottom ($b$) and top ($t$) layers. As illustrated in Fig.~\ref{fig1}(a), the moir\'e superlattices have three distinct high-symmetry locations labeled by $O$, $A$, and $B$, where the latter two positions are related by the $C_{2y}$ symmetry.

The single-particle moir\'e Hamiltonian for valence band states in $t$MoTe$_2$ has been constructed in Ref.~\onlinecite{Wu2019} and is given by,
\begin{equation}\label{ctnh0}
\begin{aligned}
&\mathcal{H}_{\tau}  =\left(\begin{array}{cc}
-\frac{\hbar^2\left(\boldsymbol{k}-\tau\boldsymbol{\bm \kappa}_{+}\right)^2}{2 m^*}+\Delta_{+}(\boldsymbol{r}) & \Delta_{\mathrm{T,\tau}}(\boldsymbol{r}) \\
\Delta_{\mathrm{T,\tau}}^{\dagger}(\boldsymbol{r}) & -\frac{\hbar^2\left(\boldsymbol{k}-\tau\bm\kappa_{-}\right)^2}{2 m^*}+\Delta_{-}(\boldsymbol{r})
\end{array}\right), \\
&\Delta_{\pm}(\boldsymbol{r}) = 2 V \sum_{j=1,3,5} \cos \left(\boldsymbol{g}_j \cdot \boldsymbol{r} \pm \psi\right), \\
&\Delta_{\mathrm{T,\tau}}(\boldsymbol{r})=w \left(1+e^{-i \tau\bm g_2 \cdot \boldsymbol{r}}+e^{-i \tau \bm g_3 \cdot \boldsymbol{r}}\right).
\end{aligned}
\end{equation} 
Here $\tau=\pm$ represents $\pm K$ valleys and also spin up ($\uparrow$) and down ($\downarrow$), since spin and valley are locked for valence band states in TMD\cite{xiao2012}. The moir\'e Hamiltonian $\mathcal{H}_\tau$ in Eq.~\eqref{ctnh0}  is represented by a $2\times2$ matrix in the layer-pseudospin space. The diagonal terms in $\mathcal{H}_\tau$ describe an electron moving in the layer-dependent potential $\Delta_\pm (\bm{r})$ with an amplitude $V$ and phase parameters $\pm \psi$, and the off-diagonal terms account for the interlayer tunneling with a strength $w$. In $\mathcal{H}_\tau$, $\bm{r}$ and $\bm{k}$ are, respectively, position and momentum operators, $m^*$ is the effective mass, the momentum shifts $\bm{\kappa}_{\pm}=\left[4\pi /(3 a_M)\right](-\sqrt{3}/2, \mp 1/2 )$ are located at corners of the moir\'e Brillouin zone, and $\bm{g}_j=\left[4\pi /(\sqrt{3} a_M)\right]\{\cos[(j-1)\pi/3], \sin[(j-1)\pi/3]\}$ for $j=1,...,6$ are the  moir\'e reciprocal lattice vectors. Here $a_M\approx a_0/\theta$ is the moir\'e period, and $a_0$ is the monolayer lattice constant. 
The moir\'e Hamiltonian $\mathcal{H}_{\tau}$ is constructed based on continuum approximation. Effects of in-plane lattice relaxation, which can be important at small twist angles ($\theta \lesssim 1^{\circ}$), are not taken into account in the Hamiltonian $\mathcal{H}_{\tau}$. The continuum approximation can also become less accurate at large twist angles ($\theta 
\gtrsim 5^{\circ}$) where the superlattices become small. Nevertheless, the Hamiltonian $\mathcal{H}_{\tau}$ qualitatively captures the evolution of electronic states as a function of $\theta$ in a convenient way.

The moir\'e Hamiltonian $\mathcal{H}_\tau$ with a fixed $\tau$ index is invariant under the $C_{3z}$ and the combined $C_{2y}\mathcal{T}$ symmetry operations, where $\mathcal{T}$ is the time-reversal symmetry. The two valleys are related by  the $C_{2y}$ and the $\mathcal{T}$ symmetries, which map $\mathcal{H}_+$ to $\mathcal{H}_-$. Moir\'e bands of $\mathcal{H}_\tau$ can be characterized by Chern numbers $C_{\tau}^{(n)}$, where $n$ is the band index (counting in descending order of energy). Because of the $\mathcal{T}$ symmetry, $C_{-}^{(n)}=-C_{+}^{(n)}$. 

A systematic study on the single-particle topological phase diagram of $\mathcal{H}_\tau$ as a function of the model parameters $(V,\psi, w)$ can be found in Ref.~\onlinecite{Pan2020}. Unless otherwise stated, we take the parameter values estimated for $t$MoTe$_2$ from Ref.~\onlinecite{Wu2019}, $a_0 = 3.472 $\AA, $m^* = 0.62 m_e$, $V = 8$ meV, $\psi = -89.6^\circ$, and $w = -8.5$ meV, where $m_e$ is the electron bare mass. The calculated moir\'e band structures are plotted in Figs.~\ref{fig1}(b) and \ref{fig1}(c) for two representative twist angles $\theta = 1.5^{\circ}$ and $2.5^{\circ}$, respectively. For twist angles $\theta \le 5^{\circ}$ studied in this work, $C_{+}^{(1)}$ of the first band in $+K$ valley is always quantized to 1. However, $(C_{+}^{(2)}, C_{+}^{(3)})$  change from $(-1, 0)$ for $\theta<\theta_1^*$ to $(1, -2)$ for $\theta>\theta_1^*$, where $\theta_1^* \approx 1.74^{\circ}$.  The topological phase transition at $\theta_1^*$ is accompanied by the gap closing between the second and the third bands at the $\bm{\gamma}$ point (i.e., the center of the moir\'e Brillouin zone). The first three moir\'e bands with a total Chern number of 0 provide a low-energy approximation of the full moir\'e band structure and are the focus of this work.

\begin{figure}[t]
\centering
\includegraphics[width=0.5\textwidth,trim=0 0 0 0,clip]{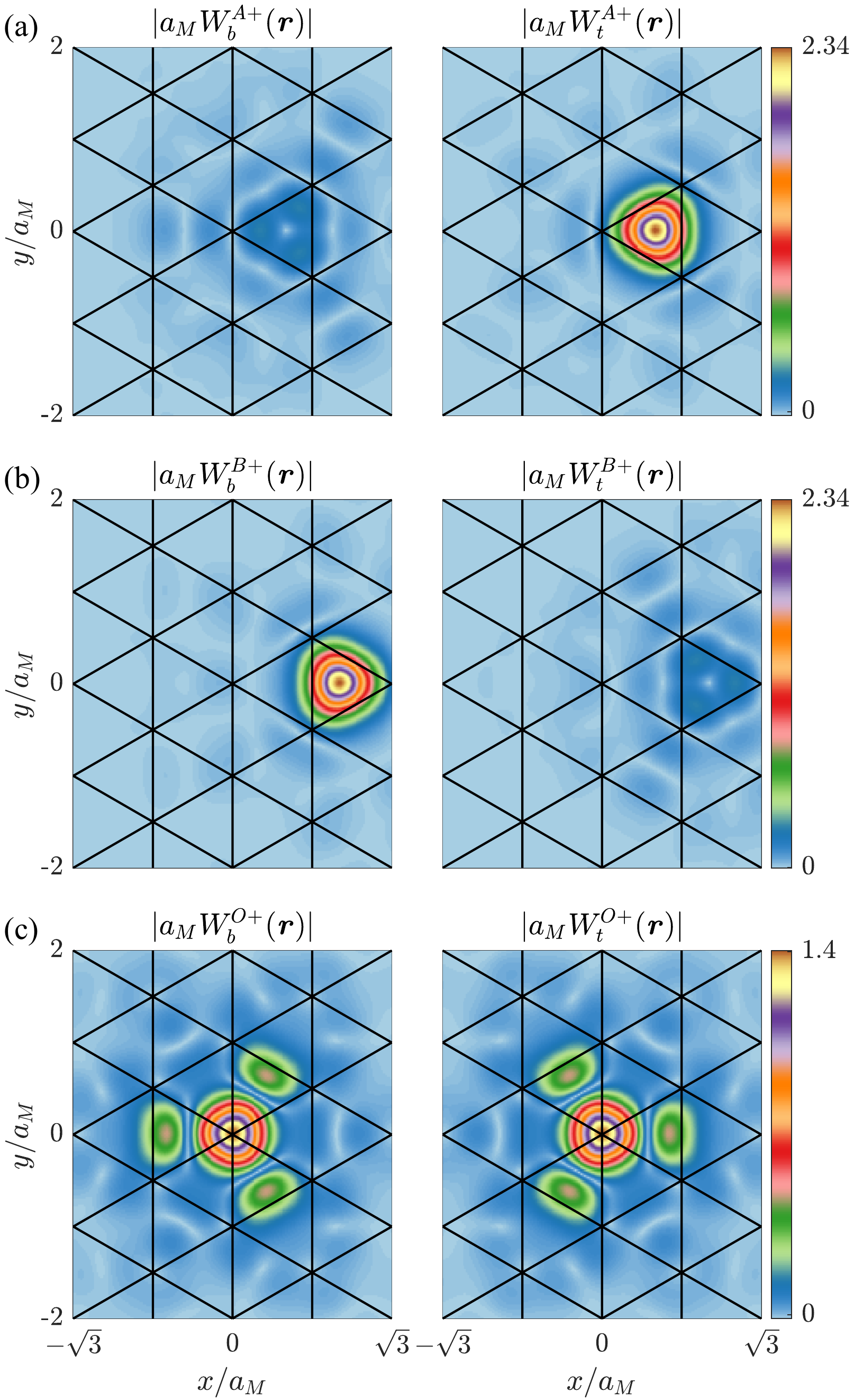}
\caption{The amplitude of each layer component of Wannier states $W^{n\tau}(\boldsymbol{r})=\left[W_b^{n\tau}(\boldsymbol{r}), W_t^{n\tau}(\boldsymbol{r})\right]^T$ for $\theta=2.5^{\circ}$ and $\tau=+$.  The black lines mark the effective triangular lattice formed by $O$ sites.}
\label{fig2}
\end{figure}

\subsection{Wannier states}

We construct Wannier states for the first three moir\'e bands so that the bands can be represented in terms of local orbitals. The center of the Wannier states can be determined by the $C_{3z}$ eigenvalues of Bloch states at the high-symmetry momenta (i.e., $\bm{\gamma}$ and $\bm{\kappa}_{\pm}$). Symmetry analysis of the moir\'e Hamiltonian and Bloch states is performed in Appendix~\ref{appa} following the approach developed in Ref.~\onlinecite{Xunjiang2022}. The $C_{3z}$ angular momenta of the Bloch states at $\bm{\gamma}$ and $\bm{\kappa}_{\pm}$ in the first three bands are labeled in Figs.~\ref{fig1}(b) and \ref{fig1}(c). The symmetry eigenvalues indicate that Wannier states for the first three bands should be centered at the $A$, $B$, and $O$ positions, respectively.  

We complement the symmetry analysis with a physical picture of the Wannier states. We first consider the case with $\theta < \theta_1^*$. In this first case, the first two bands have opposite Chern numbers $(\pm 1)$, and the third band has a Chern number of 0. Previous studies \cite{Wu2019, Devakul2021} have shown that the first two bands in one valley effectively realize a generalized Haldane model on a honeycomb lattice with two Wannier states located at the $A$ and $B$ positions, respectively.  For the third band, the $C_{3z}$ eigenvalues take the same value at $\bm \gamma$ and  $\bm\kappa_{\pm}$ ( Fig.~\ref{fig1} (b)), which implies that the corresponding Wannier states are centered at the $O$ positions. These considerations are consistent with the results obtained from the symmetry analysis. We then consider the second case with $\theta>\theta_1^*$, which is separated from the first case by a band inversion between the second and the third bands. This band inversion does not change the centers of Wannier states for the three bands as a whole.

We construct the Wannier states for $\theta<\theta_1^*$ and $\theta>\theta_1^*$ in a unified manner. 
The Wannier states in the moir\'e unit cell associated with the lattice site $\boldsymbol{R=0}$ can be formally expressed as, 
\begin{equation}\label{wnnierf}
W^{n\tau}(\boldsymbol{r})=\frac{1}{\sqrt{N}} \sum_{\boldsymbol{k}} \phi_{\boldsymbol{k}}^{n\tau}(\boldsymbol{r}), 
\end{equation}
where $\tau$ is the valley index, $n$ labels the three Wannier orbitals ($n=$1,2,3), and $N$ is the number of $\bm{k}$ points included in the summation. 
Wannier states at a generic site $\boldsymbol{R}$ are obtained through lattice translation, $W^{n\tau}_{\bm{R}}(\boldsymbol{r})=W^{n\tau}(\boldsymbol{r}-\boldsymbol{R})$.
In Eq.~\eqref{wnnierf}, $\phi_{\boldsymbol{k}}^{n\tau}(\boldsymbol{r})$ is a Bloch-like state associated with the Wannier states and related to  Bloch states of $\mathcal{H}_\tau$ by a unitary transformation,
\begin{equation}
\phi_{\boldsymbol{k}}^{n\tau}(\boldsymbol{r})=\sum_{n^{\prime}=1,2,3} U_{\boldsymbol{k} \tau}^{n^{\prime} n} \psi_{\boldsymbol{k}}^{n^{\prime}\tau}(\boldsymbol{r}),
\end{equation}
where $\psi_{\boldsymbol{k}}^{n^{\prime}\tau}(\boldsymbol{r})$ is the Bloch state of the $n^{\prime}$th moir\'e band at momentum $\boldsymbol{k}$ and  $U_{\boldsymbol{k}\tau}$ is a 3$\times$3 unitary matrix. We obtain $U_{\boldsymbol{k}\tau}$  by requiring that $\phi_{\boldsymbol{k}}^{1 \tau}$ and $\phi_{\boldsymbol{k}}^{3 \tau}$ are, respectively,  maximally polarized to the top and bottom layers. This maximum value problem is equivalent 
to find the eigenstates of the layer polarization operator $\sigma_z$ (i.e., the $z$ Pauli matrix in the layer pseudospin space) projected to the subspace spanned by $\psi_{\boldsymbol{k}}^{n\tau}$ with $n=1,2,3$, 
\begin{equation}
\Pi_{\boldsymbol{k}\tau}=\left(\begin{array}{ccc}
\left\langle\psi_{\boldsymbol{k}}^{1\tau}\left|\sigma_z\right| \psi_{\boldsymbol{k}}^{1\tau}\right\rangle & \left\langle\psi_{\boldsymbol{k}}^{1\tau}\left|\sigma_z\right| \psi_{\boldsymbol{k}}^{2\tau}\right\rangle &
\left\langle\psi_{\boldsymbol{k}}^{1\tau}\left|\sigma_z\right| \psi_{\boldsymbol{k}}^{3\tau}\right\rangle \\
\left\langle\psi_{\boldsymbol{k}}^{2\tau}\left|\sigma_z\right| \psi_{\boldsymbol{k}}^{1\tau}\right\rangle & \left\langle\psi_{\boldsymbol{k}}^{2\tau}\left|\sigma_z\right| \psi_{\boldsymbol{k}}^{2\tau}\right\rangle &
\left\langle\psi_{\boldsymbol{k}}^{2\tau}\left|\sigma_z\right| \psi_{\boldsymbol{k}}^{3\tau}\right\rangle \\
\left\langle\psi_{\boldsymbol{k}}^{3\tau}\left|\sigma_z\right| \psi_{\boldsymbol{k}}^{1\tau}\right\rangle & \left\langle\psi_{\boldsymbol{k}}^{3\tau}\left|\sigma_z\right| \psi_{\boldsymbol{k}}^{2\tau}\right\rangle &
\left\langle\psi_{\boldsymbol{k}}^{3\tau}\left|\sigma_z\right| \psi_{\boldsymbol{k}}^{3\tau}\right\rangle
\end{array}\right),
\end{equation}
where $\psi_{\boldsymbol{k}}^{n\tau}=[\psi_{\boldsymbol{k}b}^{n\tau},\psi_{\boldsymbol{k}t}^{n\tau}]^{T}$ is a two-component spinor in the layer-pseudospin space.
$U_{\boldsymbol{k}\tau}$ is then determined by,
\begin{equation}\label{vkmtx}
U_{\bm k \tau}^{\dagger} \Pi_{\bm k \tau} U_{\bm k \tau}=\left(\begin{array}{ccc}
\rho_{\bm k}^{1\tau} & 0 & 0\\
0 & \rho_{\bm k}^{2\tau} & 0\\
0 & 0 & \rho_{\bm k}^{3\tau}
\end{array}\right),
\end{equation}
where $\rho_{\bm k}^{1\tau}<\rho_{\bm k}^{2\tau}<\rho_{\bm k}^{3\tau}$. We further fix the phase of $\phi_{\boldsymbol{k}}^{n\tau}$ in the following way.  The phase of $\phi_{\boldsymbol{k}}^{1\tau}$  is chosen by requiring that its top layer component $\phi_{\boldsymbol{k}t}^{1\tau}$ is real and positive at $\bm{r}_A=a_M(1/\sqrt{3},0)$, which is the $A$ site associated with $\bm{R}=0$.   Similarly, we take the bottom layer component of $\phi_{\boldsymbol{k}}^{3\tau}$ to be real and positive at the $B$ site $\bm{r}_B=a_M(2/\sqrt{3},0)$.
For $\phi_{\boldsymbol{k}}^{2\tau}$, we first fix the gauge at $\bm{k}=\bm{\gamma}$ according to $\phi_{\boldsymbol{\gamma}b}^{2\tau}(\bm{r}_O) > 0$, where $\bm{r}_O=a_M(0,0)$; then, at a generic $\bm{k}$, we require that $\langle \phi_{\boldsymbol{\gamma}}^{2\tau}(\bm{r}_O)| \phi_{\bm{k}}^{2\tau}(\bm{r}_O)\rangle  >0$. For convenience, hereafter we denote the Bloch-like states $\phi_{\boldsymbol{k}}^{n\tau}$ with $n=1$, $2$, and $3$, respectively, as $\phi_{\boldsymbol{k}}^{A\tau}$, $\phi_{\boldsymbol{k}}^{O\tau}$, and $\phi_{\boldsymbol{k}}^{B\tau}$, and similarly for the Wannier states $W^{n\tau}$.

The Wannier states constructed for $\theta=2.5^{\circ}$ using the above methods are shown in Fig.~\ref{fig2}, where the amplitude of each layer component of $W^{n\tau}$ is plotted. The Wannier states $W^{A\tau}$ and $W^{B\tau}$ are, respectively, polarized mainly to the top and bottom layer, and transformed to each other through the $C_{2y}\mathcal{T}$ operation (plus a lattice translation). The Wannier state $W^{O\tau}$ has significant weights on both layers and is an eigenstate of the $C_{2y}\mathcal{T}$ symmetry.

A tight-binding Hamiltonian based on the Wannier states can be formally constructed in real space and then Fourier transformed to obtain the momentum-space Bloch Hamiltonian \cite{Xunjiang2022}, which has the following second-quantized form in the basis of $\phi_{\boldsymbol{k}}^{n\tau}$, 
\begin{equation}\label{3obth}
\begin{aligned}
\hat{\mathcal{H}}_0 &= \sum_{ \boldsymbol{k},\tau, n,n'}h_{\boldsymbol{k} \tau}^{n n'} c_{\boldsymbol{k} n \tau}^{\dagger} c_{\boldsymbol{k} n' \tau}, \\
h_{\boldsymbol{k} \tau} &=U_{\boldsymbol{k} \tau}^{\dagger} 
\left(\begin{array}{ccc}
E_{\bm k}^{1\tau} & 0 & 0\\
0 & E_{\bm k}^{2\tau} & 0\\
0 & 0 & E_{\bm k}^{3\tau}
\end{array}\right)
U_{\boldsymbol{k} \tau}. 
\end{aligned}
\end{equation}
Here $c_{\boldsymbol{k} n \tau}^{\dagger}$ ($c_{\boldsymbol{k} n \tau}$) is the electron creation (annihilation) operator of the Bloch-like state $\phi_{\bm{k}}^{n \tau}$, and $E_{\bm k}^{n\tau}$ is the energy of the Bloch state $\psi_{\bm{k}}^{n \tau}$ with respect to $\mathcal{H}_\tau$.

The Hamiltonian $\hat{\mathcal{H}}_0$ presents a noninteracting three-orbital model.  It faithfully captures the symmetry and topology of the first three moir\'e bands. By diagonalizing $h_{\bm{k}\tau}$, we reproduce the band structure of the first three bands and also obtain their weights projected onto the Wannier orbitals, as shown in Figs.~\ref{fig1} (b) and (c). For $\theta<\theta_1^*$, the first two moir\'e bands are mainly derived from the hybridization of $A$ and $B$ orbitals, while the third band is mainly composed of $O$ orbital. For $\theta>\theta_1^*$, the second and the third bands exchange their orbital characters at the $\bm{\gamma}$ point, which results in the band inversion.

\subsection{Interacting model}
We now construct the interacting three-orbital model.
For the charge-neutral twisted homobilayer, all valence band states are below the Fermi energy. When the system is doped with holes, it is more convenient to use the hole basis. We define the hole operator as $b_{\boldsymbol{k}n\tau}=c_{\boldsymbol{k}n\tau}^{\dagger}$. In the hole basis, the noninteracting Hamiltonian $\hat{\mathcal{H}}_0$  is equivalent to
\begin{equation}
\label{H0}
\hat{\mathcal{H}}_0=-\sum_{\boldsymbol{k}, \tau, n, n' } [h_{\boldsymbol{k}\tau}^\top]^{n n'} b_{\boldsymbol{k} n \tau}^{\dagger} b_{\boldsymbol{k} n' \tau},
\end{equation}
where a constant term is dropped.

\begin{figure*}[t]
\centering
\includegraphics[width=1.0\textwidth,trim=0 0 0 0,clip]{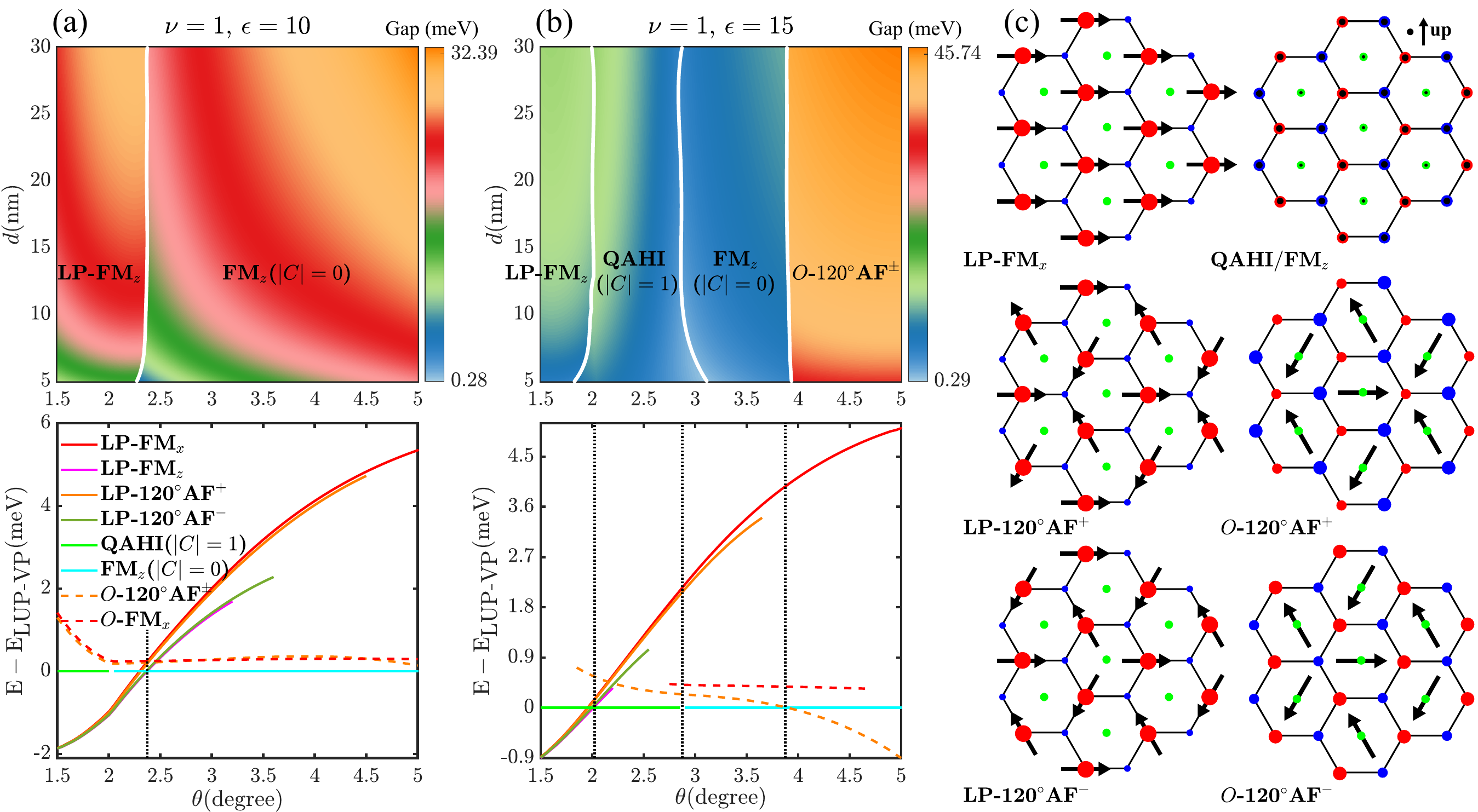}
\caption{(a), (b) Top panel: The quantum phase diagram at $\nu=1$ as a function of $\theta$ and $d$.  The color map plots the charge gap. White solid lines mark the first-order phase transitions. Bottom panel: Energy per moir\'e unit cell of competing states relative to the layer unpolarized but valley polarized (LUP-VP) states at $d=20$ nm. $\epsilon$ is 10 in (a) and 15 in (b). (c) Schematic illustration of competing states. In the $\textrm{LP-FM}_x$, $\textrm{LP-FM}_z$, $\textrm{LP-120}^{\circ}\textrm{AF}^+$ and $\textrm{LP-120}^{\circ}\textrm{AF}^-$ states, the $A$ and $B$ sublattices have unequal hole densities. In $\textrm{LP-FM}_{x}$ ($\textrm{LP-FM}_{z}$) state, $A$ (or $B$) sublattices have ferromagnetism along $x$ ($z$) direction. In $\textrm{LP-120}^{\circ}\textrm{AF}^+$ ($\textrm{LP-120}^{\circ}\textrm{AF}^-$) state, the N\'eel order has an anticlockwise (clockwise) arrangement on $A$ (or $B$) sublattices along the $y$ direction. Similar notations are used for $O$-FM$_x$ and $O$-120$^{\circ}$AF$^{\pm}$ states, of which magnetic moments are developed on the $O$ sublattice. In $\textrm{QAHI}$ and $\textrm{FM}_z$ states, the $A$ and $B$ sublattices have equal density and the magnetization is along the $z$ direction. The total Chern number is 1 in $\textrm{QAHI}$ and 0 in $\textrm{FM}_z$ states, both of which can be classified as the LUP-VP states.}
\label{fig3}
\end{figure*}

\begin{figure*}[t]
\centering
\includegraphics[width=1.0\textwidth,trim=0 0 0 0,clip]{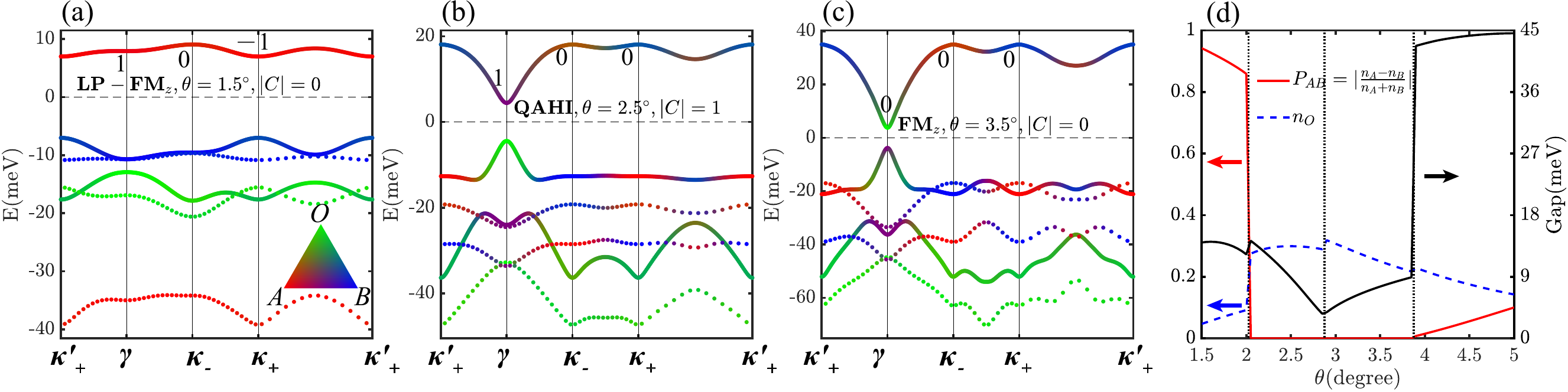}
\caption{(a)-(c) The HF band structure at $\nu=1$ in (a) the LP-FM$_z$ phase at $\theta=1.5^{\circ}$, (b) the QAHI phase at $\theta=2.5^{\circ}$, and (c) the FM$_z$ phase at $\theta=3.5^{\circ}$. The band structures are presented  in the basis defined by $c^{\dagger}_{\bm k n \tau}$ and $c_{\bm k n \tau}$ operators. The solid and dotted lines, respectively, plot bands from $+K$ and $-K$ valleys.  The color represents the weight of the three Wannier orbitals. The middle of the interaction-induced gap, marked by the black dashed line, is set to 0 in each plot. The integer numbers at $\bm{\gamma}$ and $\bm{\kappa}_{\pm}$ label the $C_{3z}$ angular momentum of the band above the Fermi energy. (d) The sublattice polarization $P_{AB}$ (red line), the hole number $n_O$ on each $O$ site (blue dashed line), and the charge gap (black line) in $\nu=1$ ground states as a function of $\theta$. $P_{AB}$ characterizes the layer polarization and is defined as $|n_A-n_B|/(n_A+n_B)$, where $n_A$ ($n_B$) is the hole number on each $A$ ($B$) site. The vertical dashed lines mark the transitions between different phases.  $d$ is 20 nm and $\epsilon$ is 15 for (a)-(d).}
\label{fig4}
\end{figure*}

The  Coulomb interaction  projected onto the three Wannier orbitals in the hole representation  is expressed as 
\begin{equation}
\label{Hint}
\begin{split}
\hat{\mathcal{H}}_\textrm{int}=\frac{1}{2} \sum V_{\bm k_1 \bm k_2 \bm k_3 \bm k_4}^{n_1 n_2 n_3 n_4}\left(\tau, \tau'\right)
b_{\bm k_1 n_1 \tau}^{+} b_{\bm k_2 n_2 \tau'}^{+} b_{\bm k_3 n_3 \tau'} b_{\bm k_4 n_4 \tau},
\end{split}
\end{equation}
where the summation is over the momentum $\bm{k}_j$ (summed over the moir\'e Brillouin zone), the orbital index $n_j$, and the valley index $\tau$.
The Coulomb matrix element is given by,
\begin{equation}
V_{\bm k_1 \bm k_2 \bm k_3 \bm k_4}^{n_1 n_2 n_3 n_4}\left(\tau, \tau'\right)=\frac{1}{\mathcal{A}} \sum_{\bm q} V_{\bm q} M_{\bm k_1 \bm k_4}^{n_1 n_4}\left(\tau, \bm q\right) M_{\bm k_2 \bm k_3}^{n_2 n_3}\left(\tau',-\bm q\right),
\end{equation}
where $\mathcal{A}$ is the system area.
The structure factor $M_{\bm k_1 \bm k_4}^{n_1 n_4}\left(\tau, \bm q\right)$ is 
\begin{equation}
M_{\bm k_1 \bm k_4}^{n_1 n_4}\left(\tau, \bm q\right)=\sum_{\ell} \int d\bm r e^{i \bm q \cdot\bm r} {[{\tilde{\phi}}_{\bm k_1 \ell}^{n_1 \tau}\left(\bm r\right)]}^{*} {\tilde{\phi}}_{\bm k_4 \ell}^{n_4 \tau}\left(\bm r\right),
\end{equation}
where $\ell$ is the layer index, and $\tilde{\phi}_{\bm k }^{n \tau}\left(r\right)=[\phi_{\bm k }^{n \tau}\left(r\right)]^*$ due to the particle-hole transformation. We use the dual-gate screened Coulomb interaction with the momentum-dependent potential $V_{\bm q}=2\pi e^2 \tanh{(|\bm q|d)}/(\epsilon |\bm q|)$, where $d$ is the gate-to-sample distance and $\epsilon$ is the dielectric constant. 
Here we neglect the effect of the vertical separation $d_0$ between the two MoTe$_2$ layers on the Coulomb potential since 
$d_0$ is much smaller than the typical interparticle distance.

By combining the noninteracting Hamiltonian $\hat{\mathcal{H}}_0$ in Eq.~\eqref{H0} with the Coulomb interaction term $\hat{\mathcal{H}}_\textrm{int}$ in Eq.~\eqref{Hint}, we obtain the full Hamiltonian of the interacting three-orbital model,
\begin{equation}
\begin{split}
\hat{\mathcal{H}}=\hat{\mathcal{H}}_0+\hat{\mathcal{H}}_\textrm{int}. 
\end{split}
\label{H3}
\end{equation}
Here we formulate the Hamiltonian $\hat{\mathcal{H}}$ in momentum space. An alternative approach would be to construct the Hamiltonian in real space by projecting the Coulomb interaction onto the Wannier states. This latter approach works well to capture onsite Hubbard interactions and offsite density-density interactions \cite{Pan2020a}, but is not suited to keep track of all relevant non-local interactions such as intersite Hund's coupling, pair hopping, and interaction-assisted hopping terms \cite{Yahui_bridging2019, Naichao_competing2021, Morales_nonlocal2022}. As shown in Fig.~\ref{fig2}, neighboring Wannier states can have significant spatial overlap, and can lead to enhanced non-local interactions that play an important role in determining the ground state properties \cite{Yahui_bridging2019, Naichao_competing2021, Morales_nonlocal2022}. Therefore, we do not use the real-space formalism. The momentum-space Hamiltonian $\hat{\mathcal{H}}$ implicitly includes all microscopic interactions, and has been widely employed in moir\'e systems such as twisted bilayer graphene \cite{Bultinck_PRX2020} to study correlated states 
 \cite{Cao2018,Cao2018a}.

\section{Phase diagram}\label{sec2}

We perform self-consistent mean-field studies of the Hamiltonian $\mathcal{H}$ using the Hartree-Fock (HF) approximation for integer filling factors $\nu=1$ and $2$ at zero temperature. Here $\nu=\frac{1}{N}\sum_{\bm k, n, \tau} b_{\bm k n \tau}^{\dagger} b_{\bm k n \tau}$ is the number of holes per moir\'e unit cell. At a given filling factor, we study multiple mean-field ansatzes that can break various symmetries of the system. The self-consistent calculation starting from different ansatz states often generates different mean-field solutions, whose energetic competitions determine the mean-field ground state. 

We calculate the phase diagram as a function of the twist angle $\theta$, the gate-to-sample distance $d$, and the dielectric constant $\epsilon$. The parameter $d$ can be experimentally varied by controlling the thickness of the encapsulating hBN layer. The screening effect by metallic gates is enhanced by reducing $d$. The dielectric constant $\epsilon$ accounts for the environmental screening from hBN as well as internal screening from remote moir\'e bands. Here we take $\epsilon$ as a phenomenological parameter and vary it theoretically to reveal the most generic phase diagram.

\subsection{$\nu=1$ phase diagram}
The quantum phase diagram at $\nu=1$ as a function of $\theta \in (1.5^\circ, 5^\circ)$ and $d \in (5 ~\text{nm}, 30 ~\text{nm})$ is shown in Figs.~\ref{fig3}(a) and \ref{fig3}(b) for $\epsilon=10$ and $15$, respectively.  We start by describing the $\theta$ dependence of the $\epsilon=15$ phase diagram.  Along a representative line cute with $d=$ 20 nm and $\epsilon=$ 15,  the ground state is in (1) the LP-FM$_z$ phase for $\theta < 2^\circ$, (2) the quantum anomalous Hall insulator (QAHI) phase for $2^\circ <\theta < 2.9^\circ$, (3) the topologically trivial FM$_z$ phase for $2.9^\circ <\theta < 3.9^\circ$, and (4) the  $O-120^{\circ}$AF$^{\pm}$ phase for $3.9^\circ <\theta < 5.0^\circ$. Here the LP-FM$_z$ phase is a multiferroic state with coexisting ferroelectricity and ferromagnetism. The ferroelectricity arises from spontaneous layer polarization (LP) induced by imbalanced hole densities on $A$ and $B$ Wannier orbitals [red line in Fig.~\ref{fig4}(d)], which is driven by the Coulomb repulsion between $A$ and $B$ sites.  The magnetism is a result of valley polarization, which leads to out-of-plane ferromagnetism (FM$_z$) since valleys are locked to out-of-plane spins in the system.
We note that the LP-FM$_z$ phase has been previously obtained in an exact-diagonalization study of $t$MoTe$_2$ for $\theta$ limited to be below $1.5^{\circ}$ \cite{Abouelkomsan_multiferroicity2022}, and multiferroic states have also been studied in moir\'eless graphene systems~\cite{Zhangfan2011,Geisenhof2021}. 
The next two phases in our phase diagram, QAHI and FM$_z$, are both valley polarized but layer unpolarized; they are distinguished by the total Chern number $C$ of the ground state, which is 1 in QAHI and 0 in FM$_z$. The QAHI and FM$_z$ phases are separated by a weakly first-order phase transition from our numerical calculations, which is manifested by a slight jump of the occupation number of $O$ orbitals across the transition [Fig.~\ref{fig4}(d)]. Since this transition is first order, the gap does not need to completely close at the transition, as shown in Fig.~\ref{fig4}(d). The topological transition is also revealed by the HF band structure in the two phases, as shown in Figs.~\ref{fig4}(b) and \ref{fig4}(c). The band (in the original basis defined by $c^{\dagger}_{\bm k n \tau}$ and $c_{\bm k n \tau}$ operators) above the Fermi level is derived from $A$ and $B$ orbitals at  $\bm{\gamma}$ and $\bm{\kappa}_{\pm}$ momenta in the QAHI phase, but is composed of $O$ orbital at $\bm{\gamma}$ point in the FM$_z$ phase. This band inversion results in the change of the Chern number. Lastly, the $O-120^{\circ}$AF$^{\pm}$ phase has the in-plane 120$^{\circ}$ antiferromagnetic (AF)  N\'eel order on the $O$ sites, and the superscript $\pm$ indicates the vector chirality of the antiferromagnetism \cite{Pan2020}, as illustrated in Fig.~\ref{fig3}c. This AF state with a given vector chirality spontaneously breaks the $C_{2y}$ symmetry, but states with opposite chiralities are related by the $C_{2y}T$ symmetry and therefore, energetically degenerate. Because of the $C_{2y}$ symmetry breaking, the $O-120^{\circ}$AF$^{\pm}$ states also have spontaneous layer polarization controlled by the vector chirality of the antiferromagnetism. This magnetoelectric coupling allows electric tuning of the vector chirality. The layer polarization is numerically confirmed by the imbalanced densities on $A$ and $B$ sites in the $O-120^{\circ}$AF$^{\pm}$ phase, as demonstrated by the red line in Fig.~\ref{fig4}(d).

\begin{figure}[b]
\centering
\includegraphics[width=1.0\columnwidth,trim=0 0 0 0,clip]{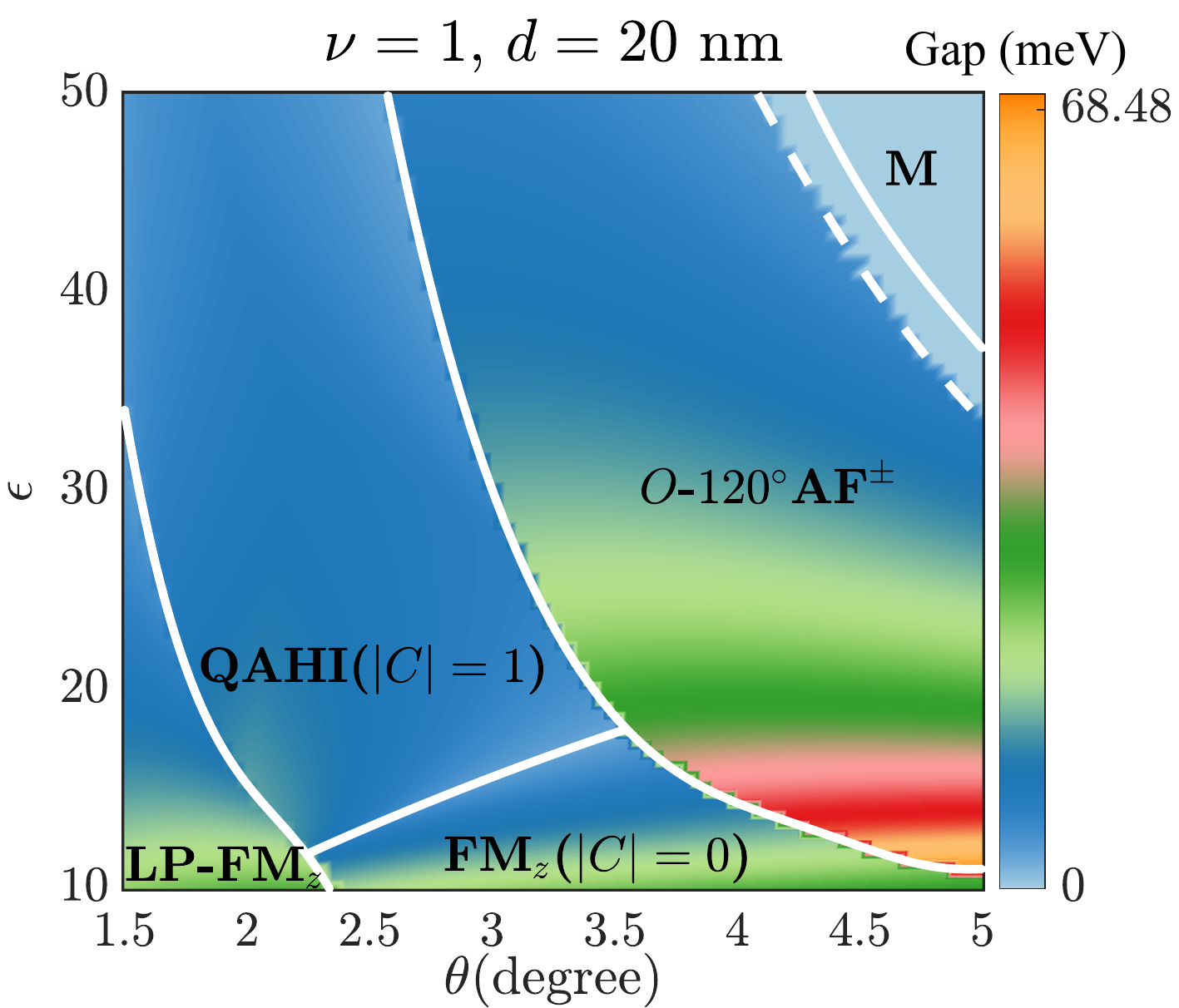}
\caption{The quantum phase diagram at $\nu=1$ and $d=20$ nm as a function of $\theta$ and $\epsilon$. The color map plots the charge gap. The symbol “$\textrm{M}$” stands for the metallic state without any symmetry breaking. The white dashed line separates gapped and gapless $O$-120$^{\circ}$AF$^{\pm}$  states.}
\label{fig4pro}
\end{figure}

\begin{figure*}[t]
\centering
\includegraphics[width=0.8\textwidth,trim=0 0 0 0,clip]{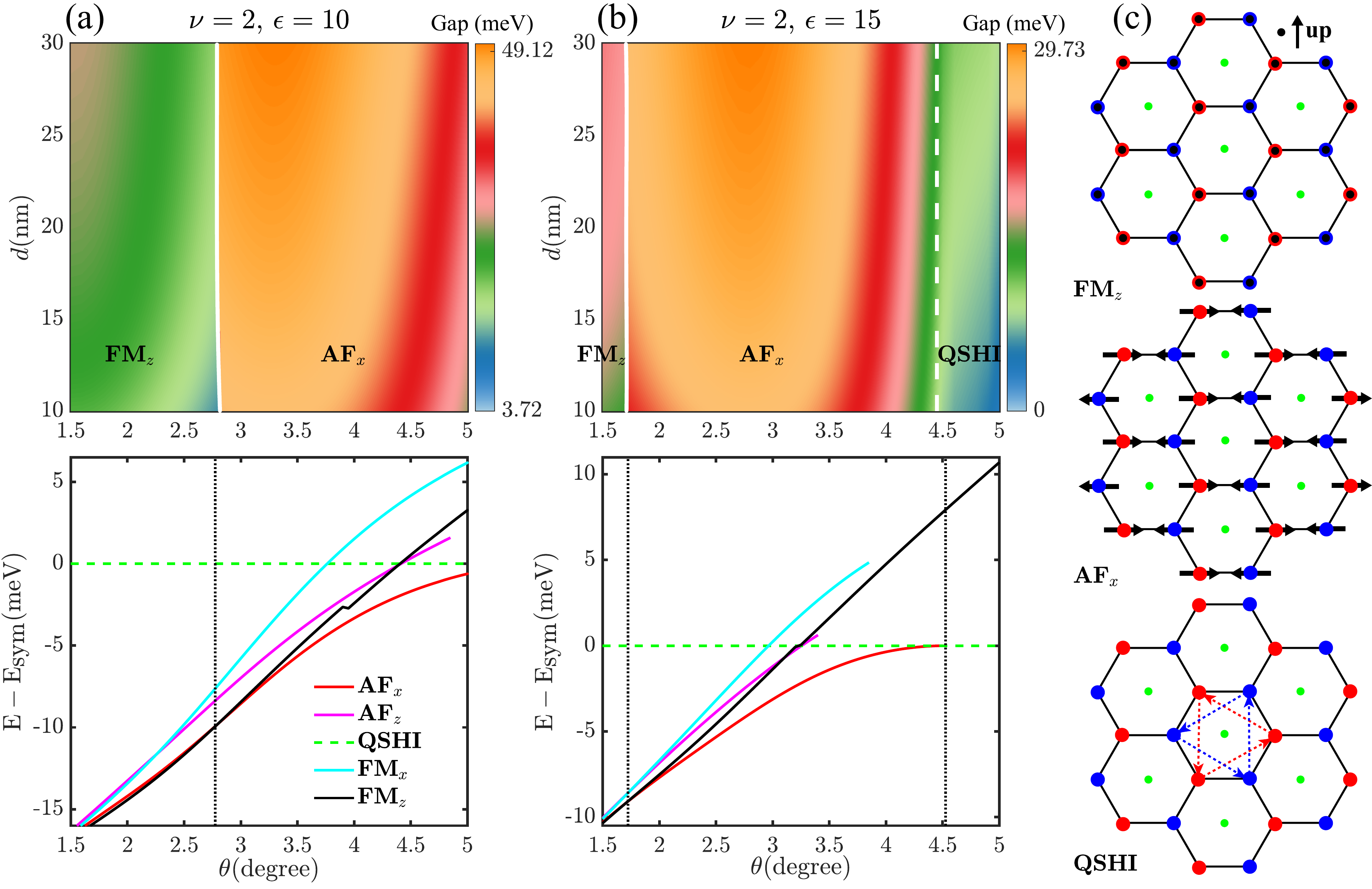}
\caption{ (a), (b) Top panel: The quantum phase diagram at $\nu=2$ as a function of $\theta$ and $d$.  The color map plots the charge gap. White solid (dashed) lines mark the first (second)-order phase transitions. Bottom panel: Energy per moir\'e unit cell of competing states relative to the symmetric (sym) state  at $d=20$ nm. $\epsilon$ is 10 in (a) and 15 in (b). (c) Schematic illustration of competing states at $\nu=2$.}
\label{fig5}
\end{figure*}

The phase diagram is determined by comparing the energies of multiple competing states, as shown in the lower panel of Figs.~\ref{fig3}(a) and \ref{fig3}(b). In addition to the states described above, we also find  LP-FM$_x$, LP-$120^{\circ}$AF$^{\pm}$, and $O$-FM$_x$ states as mean-field solutions in certain parameter regimes, although these latter states are less energetically favorable and do not appear in the ground-state phase diagram. Here LP-FM$_x$ refers to a state with spontaneous layer polarization and in-plane ferromagnetism (FM$_x$) on either $A$ or $B$ sites. LP-$120^{\circ}$AF$^{\pm}$ is also layer polarized, but has in-plane 120$^{\circ}$ AF order on either $A$ or $B$ sites; the superscript $\pm$ again indicate the vector chirality. The $O$-FM$_x$ has in-plane ferromagnetism on $O$ sites. The in-plane magnetism arises from intervalley coherent order and spontaneously breaks the valley $U(1)$ symmetry.

The phase transition between competing phases is generically first order in the $\nu=1$ phase diagram. Across the transitions, the energy of the ground state is continuous, but other quantities, such as the sublattice-resolved hole density, layer polarization, and gap, can have discontinuous jumps. The evolution of these quantities as a function of $\theta$ for the ground state is shown in Fig.~\ref{fig4}(d).  The $\theta$ dependence of the phase diagram reflects the competition between the kinetic energy and the Coulomb interaction. As the moir\'e period $a_M$ decreases by increasing $\theta$, the ratio of the characteristic Coulomb repulsion $e^2/\epsilon a_M$ to the single-particle bandwidth typically decreases, and different phases can appear.

We turn to the $\epsilon=10$ phase diagram in Fig.~\ref{fig3}(a), which includes only the LP-FM$_z$ phase for $\theta \lesssim 2.4^{\circ}$ and the FM$_z$ phase for $\theta \gtrsim
2.4^{\circ}$.  The QAHI state is a solution to the HF equations for $\theta \lesssim 2^{\circ}$, but it is energetically unfavorable compared to the LP-FM$_z$ phase in this small angle regime. The stronger Coulomb interaction for $\epsilon=10$ tends to localize carriers by repulsion. The holes are localized at $A$ or $B$ sites in the LP-FM$_z$ phase, but around $O$ sites in the FM$_z$ phase. The existence of the QAHI phase requires an intermediate strength of the Coulomb repulsion.

The evolution of the phase diagram as a function of $\epsilon$ is shown in Fig.~\ref{fig4pro}.  The region of the FM$_z$ phase in the parameter space of $(\theta, \epsilon)$ shrinks as $\epsilon$ increases and finally vanishes for $\epsilon>18$, where there is a first-order phase transition between the QAHI phase and the $O-120^{\circ}$ AF$^{\pm}$ phase. The phase boundary between the QAHI phase and the LP-FM$_z$ ($O-120^{\circ}$ AF$^{\pm}$) phase shifts to lower $\theta$ as $\epsilon$ increases. The reason is that the interaction effects become weaker with increasing $\epsilon$ but stronger with decreasing $\theta$. The $O-120^{\circ}$ AF$^{\pm}$ phase can become gapless at very large $\epsilon$ and finally give way to the metallic state without any symmetry breaking.

We now discuss the $d$ dependence of the phase diagrams. The screening from the gates mainly suppresses the Coulomb repulsion for the interaction range larger than $d$. When $d \gg a_M$, the system should have a weak dependence on $d$ since the gates only screen the long-range tail of the Coulomb repulsion in this parameter regime. Numerically, we find that the phase boundaries indeed have a relatively weak $d$ dependence for $d$ down to 10 nm, as shown in the phase diagrams of Figs.~\ref{fig3}(a) and \ref{fig3}(b). For small $d$, the phase boundary between the LP-FM$_z$ and QAHI phases in the $\epsilon=15$ phase diagram of Fig.~\ref{fig3}(b) shifts towards the LP-FM$_z$ side as $d$ decreases from 10 nm to 5 nm; the Coulomb repulsion between the nearest $A$ and $B$ sites is reduced under such small values of $d$, and the LP-FM$_z$ phase becomes less energetically favorable. A similar effect is found for the phase boundary between the FM$_z$ and QAHI phases, as shown in Fig.~\ref{fig3}(b). The numerical results indicate that the parameter regimes of the QAHI phase can be slightly enlarged by reducing $d$ down to below  $a_M$.

We present additional discussion on the $\nu=1$ phase diagram in Appendices ~\ref{appb} and \ref{appc}. In Appendix~\ref{appb}, we show that (1) the QAHI state is captured by the Kane-Mele-Hubbard model, where the Hubbard interaction combined with the spin-orbit coupling drives the out-of-plane spin polarization at $\nu=1$; (2) the transition from the LP-FM$_z$ to the QAHI state is phenomenologically described by a spinless Haldane model with nearest-neighbor repulsion;  (3) the $O-120^{\circ}$AF$^{\pm}$ state can be understood as a result of Fermi surface instability. These analyses deepen the physical understanding of the phase diagram. In Appendix~\ref{appc},  we compare results obtained, respectively, from one-band, two-band, and three-orbital models, which show that the three-orbital model is essential to capture all the competing states.

\subsection{$\nu=2$ phase diagram}

\begin{figure*}[t]
\centering
\includegraphics[width=1\textwidth]{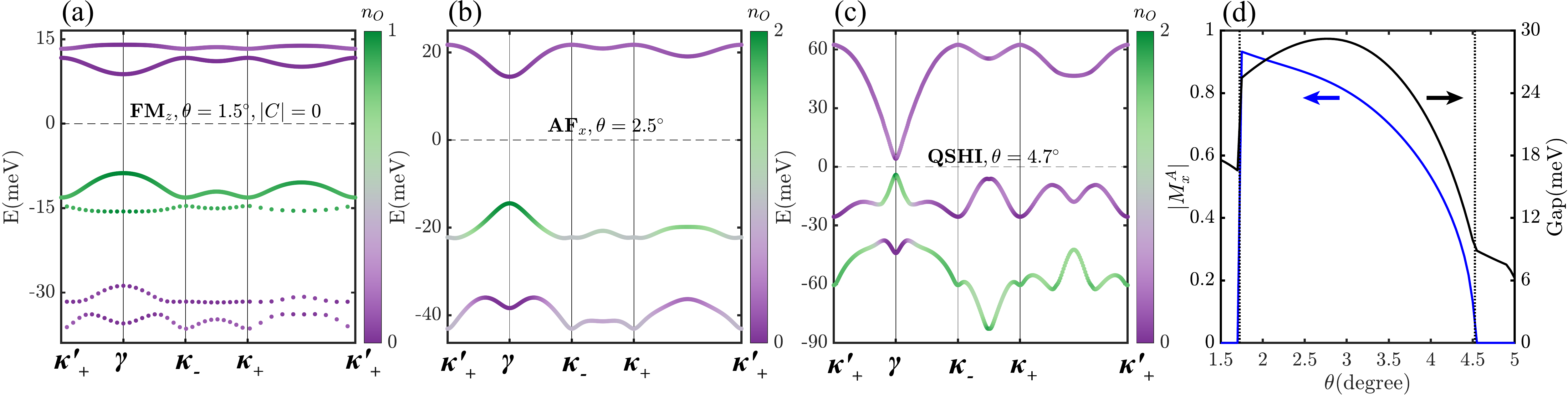}
\caption{(a)-(c) The HF band structure at $\nu=2$ in (a) the FM$_z$ phase at $\theta=1.5^{\circ}$, (b) the AF$_x$ phase at $\theta=2.5^{\circ}$, and (c) the QSHI phase at $\theta=4.7^{\circ}$. The band structures are presented  in the basis defined by $c^{\dagger}_{\bm k n \tau}$ and $c_{\bm k n \tau}$ operators. The color represents the weight of the $O$ orbital. In (a), the solid and dotted lines, respectively, plot bands from $+K$ and $-K$ valleys. In (b) and (c), each band is doubly degenerate. (d)  The charge gap (black line) and the in-plane magnetic moment $M_{x}^A$ on $A$ sublattice (blue line) in $\nu=2$ ground states as a function of $\theta$. $d$ is 20 nm and $\epsilon$ is 15 for (a)-(d).}
\label{fig6}
\end{figure*}

The quantum phase diagram at $\nu=2$ as a function of $\theta$ and $d$ is shown in Figs.~\ref{fig5}(a) and \ref{fig5}(b) for $\epsilon=10$ and 15, respectively. The phase diagram again has a weak $d$ dependence. Therefore, we focus on the $\theta$ dependence of the phase diagram. The ground state along the line cut with $d=$ 20 nm and $\epsilon=$ 15 is in (1) the FM$_z$ phase for $\theta<1.7^\circ$, (2) the AF$_x$ phase for $1.7^\circ<\theta<4.5^\circ$, and (3) the quantum spin Hall insulator (QSHI) phase for $\theta>4.5^\circ$. The phase diagram is obtained by comparing the energy of multiple competing states, including FM$_z$, FM$_x$, AF$_x$, AF$_z$, and symmetric states.  All these states have equal hole densities on $A$ and $B$ sublattices. The symmetric state does not break any symmetry of the system and always realizes the QSHI state for the $\theta$ range $(1.5^{\circ}, 5^{\circ})$ under consideration. In the FM$_z$ (FM$_x$) state, the $A$ and $B$ sublattices have ferromagnetic moments along $z$ ($x$) direction. The AF$_z$ (AF$_x$) state has collinear antiferromagnetic N\'eel order with opposite magnetic moments on $A$ and $B$ sublattices along $z$ ($x$) direction. The HF band structures for the FM$_x$, AF$_x$, and QSHI phases are, respectively, shown in Figs.~\ref{fig6}(a), \ref{fig6}(b), and \ref{fig6}(c), where the color encodes the weight on the $O$ orbital. The results show that the holes at $\nu=2$ are primarily doped to the $A$ and $B$ sites.

In the small $\theta$ regime of $\theta<1.7^\circ$, the energy of the FM$_z$ state is lower than that of the AF$_x$ state only by a very small margin, as shown by the bottom panel of Fig.~\ref{fig5}(b). The transition between the FM$_z$ and AF$_x$ phases at $\theta \approx 1.7^{\circ}$ is first order. By contrast, the transition between the AF$_x$ and QSHI phases at $\theta \approx 4.5^{\circ}$ is second order. This is manifested by the continuous suppression of the N\'eel order as $\theta$ approaches 4.5$^{\circ}$ from the AF$_x$ phase [blue line in Fig.~\ref{fig6}(d)]. This phase transition between the AF$_x$ and QSHI phases is reminiscent of the same transition in the Kane-Mele-Hubbard model at a critical onsite Hubbard interaction \cite{Hohenadler2011Corr}. While our realistic model differs from the Kane-Mele-Hubbard model due to complications such as the additional $O$ orbital and the long-range Coulomb interaction, the latter model provides important intuitions in understanding the phase diagram.  Mean-field theory underestimates the critical Hubbard repulsion in the Kane-Mele-Hubbard model compared to that obtained from  quantum Monte Carlo simulation \cite{Hohenadler2012quant}. It is likely that our mean-field phase diagram also overestimates the tendency towards the formation of the N\'eel order, and the QSHI phase could appear in a larger parameter space.

In the QSHI phase, the occupied bands in opposite valleys have opposite Chern numbers of $\pm 1$. Since the two valleys also act as spin up and down that are related by the $\mathcal{T}$ symmetry, the QSHI state is characterized by nontrivial $Z_2$ topological invariant. In the single-particle band structure of the noninteracting moir\'e Hamiltonian, the first and the second band overlap in energy for $\theta>3.1^{\circ}$, and the system would then be a metallic state at $\nu=2$ without taking into account of interaction effects. However, the QSHI phase of the interacting model has a full charge gap as indicated by the HF band structure shown in Fig.~\ref{fig6}(c). Thus, the Coulomb repulsion is crucial in opening up the charge gap in the QSHI phase.

\begin{figure}[b]
\centering
\includegraphics[width=1.0\columnwidth,trim=0 0 0 0,clip]{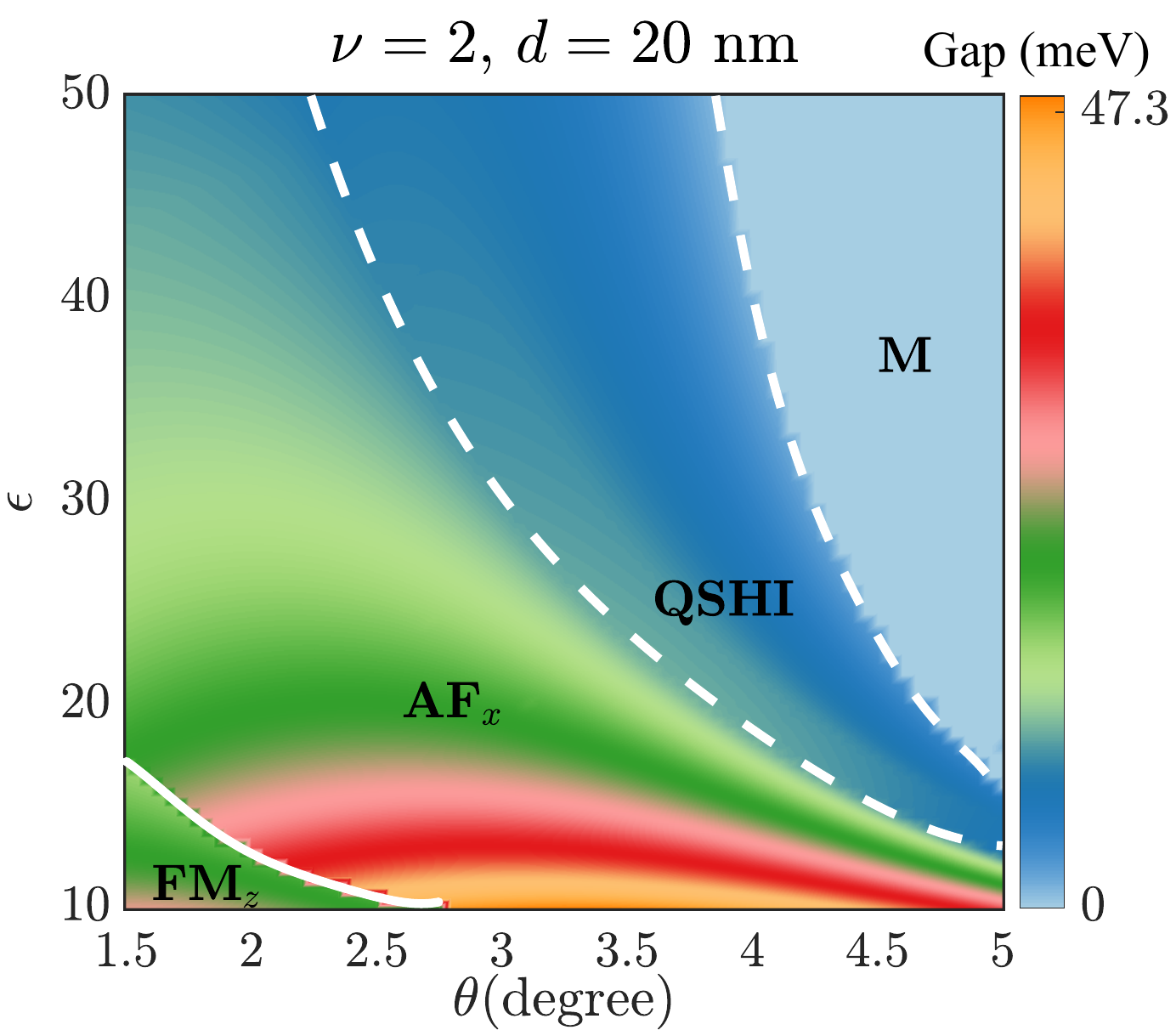}
\caption{The quantum phase diagram at $\nu=2$ and $d=20$ nm as a function of $\theta$ and $\epsilon$. The color map plots the charge gap. The symbol “$\textrm{M}$” stands for metallic state.}
\label{fig6pro}
\end{figure}

When $\epsilon$ is reduced to 10, the corresponding phase diagram in Fig.~\ref{fig5}(a) only has the FM$_z$ and AF$_x$ phases for $\theta \in (1.5^\circ, 5^\circ)$. The stronger Coulomb interaction for $\epsilon=10$ enhances the antiferromagnetic order and therefore, the QSHI state no longer appears in the phase diagram of Fig.~\ref{fig5}(a). The existence of the QSHI state requires an intermediate interaction strength, which should be strong enough to open a full charge gap but weak enough to avoid the instability toward magnetism.
This is clearly shown by the phase diagram as a function of $\theta$ and $\epsilon$  plotted in Fig.~\ref{fig6pro}.

\section{Discussion and Conclusion}\label{sec3}
In summary, we present a systematic study of the quantum phase diagrams of $t$MoTe$_2$ that result from the interplay between band topology and many-body interaction. At the single-particle level, we construct a three-orbital model that can be viewed as a generalization of the two-orbital Kane-Mele model. The three-orbital model faithfully captures the band dispersion, symmetry, and topology of the low-energy moir\'e bands for a large range of $\theta$. For the interacting physics, we obtain the quantum phase diagrams based on comprehensive many-body calculations by comparing the energy of multiple competing states.  At $\nu=1$, we find a cascade of phase transitions as $\theta$ increases. In the small $\theta$ regime, the system is in a layer-polarized phase with spontaneous imbalanced hole densities on $A$ and $B$ sites. The  QAHI phase appears in an intermediate range of $\theta$ and is superseded by topologically trivial phases (i.e., FM$_z$ and $O-120^{\circ}$AF$^{\pm}$ phases) in the large $\theta$ regime. At $\nu=2$, the QSHI phase can appear in the large $\theta$ regime, but undergo an instability towards an antiferromagnetic state below a critical $\theta$.

The obtained phase diagrams serve as a guide for further experimental and theoretical studies, but the results should be taken to be qualitative instead of quantitative. There are several theoretical difficulties in establishing truly quantitative phase diagrams. First, parameters in the moir\'e Hamiltonian  are subjected to numerical uncertainties. We take the parameter values from Ref.~\cite{Wu2019}, where parameters were obtained by fitting to the density-functional-theory (DFT) band structures of the untwisted homobilayer at different high-symmetry stackings. The DFT calculations are likely not accurate enough to pin down the exact values of small quantities such as $V$ and $w$ that are on the order of 10 meV. Second, mean-field theory can capture the quantum phase diagram qualitatively but may overestimate the tendency towards symmetry-breaking states. In spite of these quantitative issues that are generic challenges for moir\'e systems, we expect that our study captures the qualitative features of the phase diagrams.

\begin{figure}[t]
\centering
\includegraphics[width=0.45\textwidth,trim=3 0 0 0,clip]{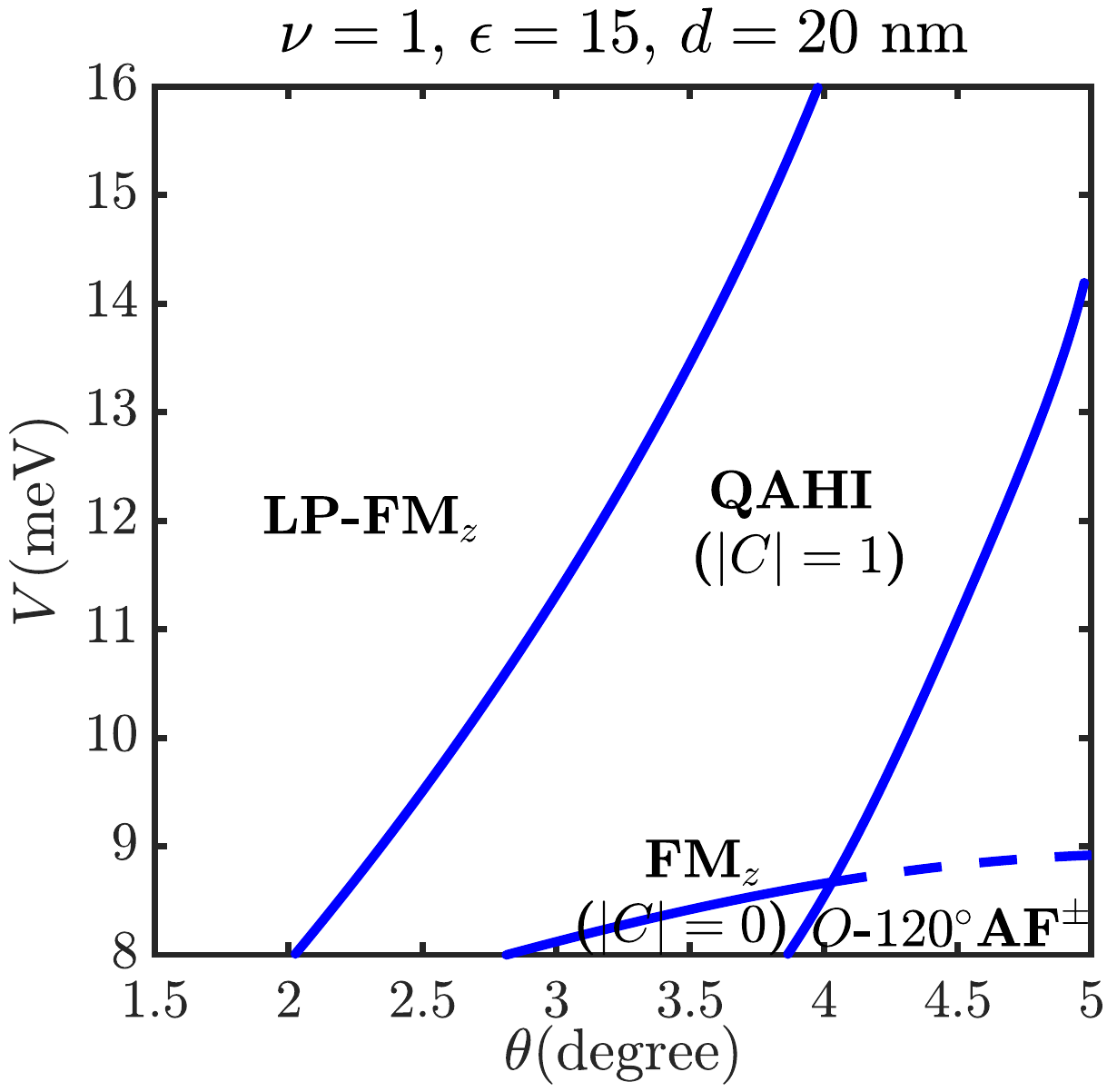}
\caption{Quantum phase diagram at $\nu=1$ as a function of $\theta$ and $V$ for $d=20$ nm and $\epsilon = 15$. Different phases are separated by blue solid lines. The blue dashed line is a continuation of the phase boundary between the FM$_z$ and QAHI phases if the $O-120^{\circ}$AF$^{\pm}$ phase is fully suppressed by an out-of-plane magnetic field.}
\label{fig7}
\end{figure}

Recent experiments on $t$MoTe$_2$ reported optical signatures of a ferromagnetic correlated insulator in a $\theta=3.9^{\circ}$ sample \cite{Xiaodong2023a} and a quantum anomalous Hall state with Chern number of 1 in another $\theta=3.7^{\circ}$ sample \cite{Xiaodong2023b}, where the hole filling factor $\nu$ is 1 in both cases. The topological character of the $\nu=1$ ferromagnetic insulator in the $\theta=3.9^{\circ}$ sample was not reported \cite{Xiaodong2023a}. An independent experiment 
reported the quantum anomalous Hall state in $t$MoTe$_2$ with $\theta=3.4^{\circ}$  at $\nu=1$ \cite{Yihang2023_integer}.
Our $\nu=1$ phase diagram featuring both the topological ferromagnetic phase (i.e., the QAHI phase ) and the topologically trivial ferromagnetic phase (i.e., the FM$_z$ phase) is qualitatively consistent with these experimental findings.  In the phase diagram of Fig.~\ref{fig3}(b), the QAHI phase appears for $\theta$ in the range of about $(2^{\circ}, 2.9^{\circ})$, which is below the experimental twist angle $3.4^{\circ}$. This discrepancy can be remedied by slightly adjusting the parameters used in the moir\'e Hamiltonian. In Fig.~\ref{fig7}, we present the $\nu=1$ phase diagram as a function of $\theta$ and $V$, where $V$ is the moir\'e potential amplitude as defined in Eq.~\eqref{ctnh0}. In the calculation of Fig.~\ref{fig7}, we keep other parameters (except $\theta$ and $V$) of the moir\'e Hamiltonian in Eq.~\eqref{ctnh0} fixed, and take $\epsilon=15$ and $d=20$ nm.   As shown in Fig.~\ref{fig7}, the $\theta$ range for the QAHI phase is shifted to $(2.35^{\circ}, 4.05^{\circ})$ if $V$ is increased to 9 meV, which leads to better consistency with the experimental finding \cite{Xiaodong2023b}. Increasing the intralayer moir\'e potential amplitude $V$ has two effects. First, a deeper moir\'e potential leads to stronger spatial confinement of $A$ and $B$ orbitals, and therefore, weaker hopping parameters between $A$ and $B$ sites. The doped holes then have a stronger tendency to be localized by Coulomb repulsion, which shifts the phase boundary between the LP-FM$_z$ and QAHI phase to larger twist angles. Second, the deeper intralayer moir\'e potential also increases the single-particle energy difference between $A$/$B$ orbital and  $O$ orbital, and effectively pushes the $O$ orbital away from the Fermi level. Therefore, the topologically trivial FM$_z$ phase gradually shrinks as $V$ increases and finally disappears at  $V \approx 8.7$ meV. For $V > 8.7$ meV, there is a first-order phase transition between the QAHI phase and the $O-120^{\circ}$AF$^{\pm}$ phase, where the corresponding phase boundary also shifts towards larger twist angles as $V$ increases.  We note that the moir\'e potential could in principle be tuned by pressure \cite{Yankowitz2019,Morales_pressure2023}, but the exact dependence of the model parameters and the phase diagrams on the pressure requires future investigations.

We comment on the implication of our results for $t$WSe$_2$. Early transport studies of $t$WSe$_2$ reported topologically trivial correlated insulator at $\nu=1$ in the twist angle range from $4^{\circ}$ to $5.5^{\circ}$ \cite{Wang2020,Ghiotto2021_Quantum}. Recent local electronic compressibility measurements of $t$WSe$_2$ at $\nu=1$ revealed QAHI state in a narrow twist angle range between $1.2^{\circ}$ and $1.25^{\circ}$, but topologically trivial correlated insulator for $\theta$ either below or above this range \cite{Foutty_mapping2023}. Qualitatively, our $\nu=1$ phase diagram with the QAHI phase appearing in an intermediate range of $\theta$ is consistent with these experimental observations. The $O-120^{\circ}$AF$^{\pm}$ phase and the LP phase can capture the topologically trivial correlated insulators observed at large and small twist angles, respectively. Quantitatively, one peculiar experimental finding is that the QAHI state in $t$WSe$_2$ only appears at a very narrow twist angle range \cite{Foutty_mapping2023}, which calls for further investigation.

Signatures of fractional quantum anomalous Hall (FQAH) states at fractional filling factors were also reported in $t$MoTe$_2$ \cite{Xiaodong2023b, Yihang2023_integer}.  Exact diagonalization studies of the FQAH states have recently been performed, where the calculations were done in the truncated Hilbert space projected to the first moir\'e valence band \cite{wangchong_fractional, fuliang_fractional}. Our study highlights the importance of including multiple bands in the many-body problem. For example, the fully valley polarized state in the truncated Hilbert space projected to the first moir\'e valence bands is always a quantum anomalous Hall state at $\nu=1$, but this is not the case in the three-orbital model (see more discussion in Appendix~\ref{appc}). It is desirable to examine the energy competition between the FQAH state and other candidate states (e.g., charge density wave state) at fractional filling factors in the multi-band model using techniques beyond the mean-field theory. 

FQAH states are analogous to fractional quantum Hall states formed in Landau levels. However, there is a crucial difference between Landau levels and topological moir\'e bands in $t$MoTe$_2$, since the latter respects time-reversal symmetry. The FQAH states realized in $t$MoTe$_2$ spontaneously break the time-reversal symmetry. Time-reversal symmetric fractional quantum spin Hall insulators \cite{Levin2009,Qi2011}, as another distinct type of fractionalized states, may also emerge in $t$MoTe$_2$ and $t$WSe$_2$.

The obtained quantum phases can have interesting responses to external fields. For example, an out-of-plane electric displacement field can drive the QAHI state at $\nu=1$ into a topologically trivial layer polarized state. An out-of-plane magnetic field turns the $\nu=1$ $O-120^{\circ}$AF$^{\pm}$ state into a canted antiferromagnetic state, and finally to a ferromagnetic state for magnetic field above a critical value, where the ferromagnetic state can be in the QAHI phase, as shown in Fig.~\ref{fig7}.
The exploration of the rich quantum states in $t$MoTe$_2$ is just at its beginning stage.

\section{ACKNOWLEDGMENTS}
This work is supported by National Key Research and Development Program of China (Grants No. 2021YFA1401300 and No. 2022YFA1402401), National Natural Science Foundation of China (Grant No. 12274333), and start-up funding of Wuhan University. W.-X. Q. is also supported by Postdoctoral Innovation Research Position of Hubei Province (Grant No. 211000128). The numerical calculations in this paper have been done on the supercomputing system in the Supercomputing Center of Wuhan University.

 \vspace{\baselineskip}
\textit{Note added.} Fractionally quantized anomalous Hall effect has recently been observed in  $t$MoTe$_2$ through transport measurement \cite{Park2023b,Xu2023}.

\appendix

\section{$C_{3z}$ symmetry eigenvalues}\label{appa}

The single-particle wave function can be viewed as a direct product of the spatial envelope wave function $\psi_{\bm{k}}^{n \tau} (\bm{r})$, the atomic state $|d_\tau\rangle$, and the spin state $|s_\tau\rangle$, where $|d_\tau\rangle=|d_{x^2-y^2}\rangle + i \tau |d_{xy}\rangle$ stands for the dominant atomic orbital, and $|s_\tau\rangle$ represents spin up and down, respectively, for $\tau=+$ and $-$. The $C_{3z}$ symmetry acts separately on the three parts. For the internal degrees of freedom, the $C_{3z}$ symmetry eigenvalue is given by,
\begin{equation}
    C_{3z} |d_\tau\rangle \otimes |s_\tau\rangle = e^{i 2\pi l_{\tau} /3} |d_\tau\rangle \otimes |s_\tau\rangle,
\end{equation}
where $l_{\tau} = -\tau/2$ is the angular momentum contributed by the spin and atomic orbital. 

We now study the $C_{3z}$ symmetry property of the spatial wave function using the approach developed in Ref.~\onlinecite{Xunjiang2022}. 
We first apply the following unitary transformation to $\mathcal{H}_{\tau}$ to make the $C_{3z}$ symmetry transparent,
\begin{equation}
\begin{aligned}
& \tilde{\mathcal{H}}_{\tau}(\bm r)\equiv \Lambda_{\tau}(\bm r)\mathcal{H}_{\tau}(\bm r)\Lambda_{\tau}^{-1}(\bm r),\\
&\Lambda_{\tau}(\bm r)=\begin{pmatrix} e^{-i \tau\bm \kappa_{+} \bm \cdot\bm r}& 0\\
0& e^{-i \tau\bm \kappa_{-} \bm \cdot\bm r}\end{pmatrix},\\
&\tilde{\mathcal{H}}_{\tau}(\bm r)=\begin{pmatrix}
-\frac{\hbar^2\boldsymbol{k}^2}{2 m^*}+\Delta_{+}(\bm r)&\tilde{\Delta}_{\mathrm{T,\tau}}(\bm r)\\
\tilde{\Delta}_{\mathrm{T,\tau}}^{\dagger}(\bm r)&-\frac{\hbar^2\boldsymbol{k}^2}{2 m^*}+\Delta_{-}(\bm r)
\end{pmatrix},
\label{C31}
\end{aligned}
\end{equation}
where $\tilde{\Delta}_{\mathrm{T,\tau}}(\bm r)=w(e^{i\tau\bm {q_1\cdot r}}+e^{i\tau\bm {q_2}\bm \cdot\bm r}+e^{i\tau\bm q_3\bm \cdot\bm r})$. Here $\bm q_1=\bm\kappa_{-}-\bm \kappa_{+},\bm q_2=\hat{R}_{3}\bm q_1$,  $\bm q_3=\hat{R}_{3}\bm q_2$, and $\hat{R}_{3}$ is the real-space anticlockwise rotation by $2\pi/3$. The new Hamiltonian $\tilde{\mathcal{H}}_{\tau}(\bm r)$ is clearly invariant under the threefold rotation of $\bm r$ since $\tilde{\mathcal{H}}_{\tau}(\hat{R}_{3} \bm r) = \tilde{\mathcal{H}}_{\tau}(\bm r)$.

The $C_{3z}$ symmetry of the original Hamiltonian $\mathcal{H}_\tau$ can be represented as $\hat{C}_{3z,\tau}=\Lambda_{\tau}^{-1}(\bm r) \hat{R}_3 \Lambda_{\tau}(\bm r)$, which leads to the symmetry identity $\hat{C}_{3z,\tau} \mathcal{H}_\tau(\bm{r}) \hat{C}_{3z,\tau}^{-1} = \mathcal{H}_\tau(\bm{r})$. The $C_{3z}$ symmetry acts on the wave function in the following way,
\begin{equation}
\begin{aligned}
\hat{C}_{3z,\tau}\psi_{\bm k}^{n\tau}(\bm r)
=\Lambda_{\tau}^{-1}(\bm r) \hat{R}_3 \tilde{\psi}_{\bm{k}}^{n \tau}(\bm{r}),
\end{aligned}
\label{C32}
\end{equation}
where $\tilde{\psi}_{\bm{k}}^{n \tau}(\bm{r})= \Lambda_{\tau}(\bm{r}) \psi_{\bm{k}}^{n \tau}(\bm{r}) $.

At the three high-symmetry momenta $\bm{\gamma}$ and $\bm{\kappa}_{\pm}$, the Bloch wave function is an eigenstate of the $C_{3z}$ symmetry,
\begin{equation}
\hat{C}_{3z,\tau}{\psi}_{\bm k}^{n\tau}(\bm r)=e^{i2\pi L_{\bm k}^{n\tau}/3}{\psi}_{\bm k}^{n\tau}(\bm r),
\label{C36}
\end{equation}
where $L_{\bm k}^{n\tau}$ is the $C_{3z}$ angular momentum of the spatial wave function and defined modulo 3. Using Eqs.~\eqref{C32} and \eqref{C36}, we find that 
\begin{equation}
\begin{aligned}
\tilde{\psi}_{\bm k}^{n\tau}(\hat{R}_3\bm r)=e^{i2\pi L_{\bm k}^{n\tau}/3}\tilde{\psi}_{\bm k}^{n\tau}(\bm r).
\label{C37}
\end{aligned}
\end{equation}
Therefore, $L_{\bm k}^{n\tau}$ can be determined by numerically comparing  $\tilde{\psi}_{\bm k}^{n\tau}(\hat{R}_3\bm r)$ and $\tilde{\psi}_{\bm k}^{n\tau}(\bm r)$. The values of $L_{\bm k}^{n\tau}$ for the first three bands at the three high-symmetry momenta are labeled in Figs.~\ref{fig1}(b) and \ref{fig1}(c). 

In a system with $C_{3z}$ symmetry \cite{Fang2012}, the Chern number of a band is determined by the sum of the $C_{3z}$ angular momentum up to modulo 3, $[(L_{\bm {\gamma}}^{n\tau}+L_{\bm {\kappa_+}}^{n\tau}+L_{\bm {\kappa_-}}^{n\tau})-C_{\tau}^{(n)}]$ mod $3 =0 $. The calculated Chern numbers and $C_{3z}$ angular momentum are consistent with this condition.

\section{Physical understanding}\label{appb}
In this Appendix, we present analyses that help to deepen understanding of the $\nu=1$ phase diagram. As explained in the following, we show that (1) the QAHI state is captured by the Kane-Mele-Hubbard model, (2) the transition from the LP-FM$_z$ to the QAHI state is phenomenologically described by the Haldane-$V$ model (i.e., spinless Haldane model with nearest-neighbor repulsion $V$), and (3) the $O-120^{\circ}$AF$^{\pm}$ state can arise from Fermi surface instability. We emphasize that these analyses should be taken to be qualitative instead of quantitative. The quantitative energy competition between different phases depends crucially on microscopic details and can only be obtained from detailed numerical calculations.

\subsection{Kane-Mele-Hubbard model}
We first consider the Kane-Mele-Hubbard model, which can be constructed by retaining only the $A$ and $B$ orbitals and neglecting remote hoppings and repulsions. The model is given by,
\begin{equation}
\begin{aligned}
\hat{\mathcal{H}}_{\textrm{KMH}}&=\hat{\mathcal{H}}_{\textrm{KM}}+\hat{\mathcal{H}}_U, \\
\hat{\mathcal{H}}_{\textrm{KM}}&=\sum_{\langle i j\rangle s} t_0 b_{i s}^{+} b_{j s}+\sum_{\langle\langle i j\rangle\rangle s} t_{i j}^s b_{i s}^{+} b_{j s}\\
&=\sum_{\bm{k}} b_{\boldsymbol{k}}^{+}\left[F_0 \sigma_0 s_0+F_x \sigma_x s_0+F_y \sigma_y s_0+F_z \sigma_z s_z\right] b_{\boldsymbol{k}}, \\
\hat{\mathcal{H}}_U&=U \sum_i \hat{n}_{i \uparrow} \hat{n}_{i \downarrow},
\end{aligned}
\end{equation}
where $\hat{\mathcal{H}}_{\textrm{KM}}$ is the Kane-Mele model defined on a honeycomb lattice, the fermion operators $b_{i s}^{+}$ and $b_{i s}$ are defined in the hole basis, the subscript $s$ is the spin index (locked to the valley index in $t$MoTe$_2$), $t_0$ is the nearest neighbor hopping parameter, and $t_{i j}^s$ describes the second nearest neighbor hopping with spin and sublattice-dependent flux. Here $b_{\boldsymbol{k}}=\{b_{\boldsymbol{k}A\uparrow},b_{\boldsymbol{k}A\downarrow},b_{\boldsymbol{k}B\uparrow},b_{\boldsymbol{k}B\downarrow}\}^{\mathrm{T}}$, where the subscripts $A$ and $B$ denote the sublattices in the honeycomb lattice. $\sigma_{x,y,z}$ and $s_{x,y,z}$ are, respectively, Pauli matrices in the sublattice and spin spaces, and $F_{0,x,y,z}$ are real functions of the momentum $\boldsymbol{k}$.

Using Hartree-Fock decomposition, the Hubbard term $\hat{\mathcal{H}}_U$ can be approximated as
\begin{equation}
\hat{\mathcal{H}}_U \approx-\frac{N U}{4} \sum_{\ell}\left[n_{\ell}^2-\boldsymbol{m}_{\ell}^2\right]+\frac{U}{2} \sum_{\boldsymbol{k} \ell} b_{\boldsymbol{k} \ell}^{+}\left[n_{\ell} s_0-\boldsymbol{m}_{\ell} \cdot \boldsymbol{s}\right] b_{\boldsymbol{k} \ell},
\end{equation}
where the lattice translational symmetry is assumed to be preserved, $N$ is the number of unit cells, $n_{\ell}=\left\langle{b}_{\ell}^{+}{b}_{\ell}\right\rangle$, $\boldsymbol{m}_{\ell}=\left\langle b_{\ell}^{+} \boldsymbol{s} b_{\ell}\right\rangle$, $b_{\ell}=\{b_{\ell\uparrow},b_{\ell\downarrow}\}^{\mathrm{T}}$, and $\ell$ is the sublattice index.

By diagonalizing the mean-field Hamiltonian, we obtain the total energy at $\nu=1$ as follows,
\begin{center}
\begin{equation}
\begin{aligned}
&E_{tot}=\frac{N U}{8}+\frac{2 N M^2}{U}+\sum_{\bm k} F_0-\sum_{\bm{k}} \sqrt{F^2+M^2+\Phi}, \\
&\Phi=F_z M \left(\cos \theta_A-\cos \theta_B\right) + \sqrt{2 M^2 \left|F_{\|}\right|^2\Phi_1 +M^2 F_z^2\Phi_2},\\
&\Phi_1=1+\cos \theta_A \cos \theta_B+\sin \theta_A \sin \theta_B \cos \left(\phi_A-\phi_B\right),\\
&\Phi_2=\left(\cos \theta_A+\cos \theta_B\right)^2,
\end{aligned}
\end{equation}
\end{center}
where $\bm{M}_{\ell}=-U\bm{m}_{\ell}/2$, $F^2=|F_{\|}|^2+F^2_z$, $F_{\|}=F_x-iF_y$, and $\bm{M}_{\ell}$ is parameterized as $M_{\ell}(\sin\theta_{\ell}\cos\phi_{\ell},\sin\theta_{\ell}\sin\phi_{\ell},\cos\theta_{\ell})$. In deriving the total energy, we assume that $n_A=n_B$ and $M_A=M_B=M$. It is obvious that $E_{tot}$ is minimized when $\phi_A=\phi_B$. By further taking $\partial E_{tot} / \partial \theta_{\ell}=0$, we find two types of solutions, (1) $\theta_A=\theta_B=0$ and (2) $\theta_A=0$, $\theta_B=\pi$. It can be shown analytically that $E_{tot}(\theta_A=\theta_B=0)<E_{tot}(\theta_A=0,\theta_B=\pi)$. Therefore, the Kane-Mele-Hubbard model supports the out-of-plane spin polarized state at $\nu=1$, which exactly gives rise to the valley polarized QAHI state in $t$MoTe$_2$. From the expression of $E_{tot}$, we see that the spin-dependent term $F_z \sigma_z s_z$ in the Kane-Mele model leads to Ising anisotropy and stabilizes the QAHI state.

\begin{figure}[t]
\centering
\includegraphics[width=1.0\columnwidth,trim=3 0 0 0,clip]{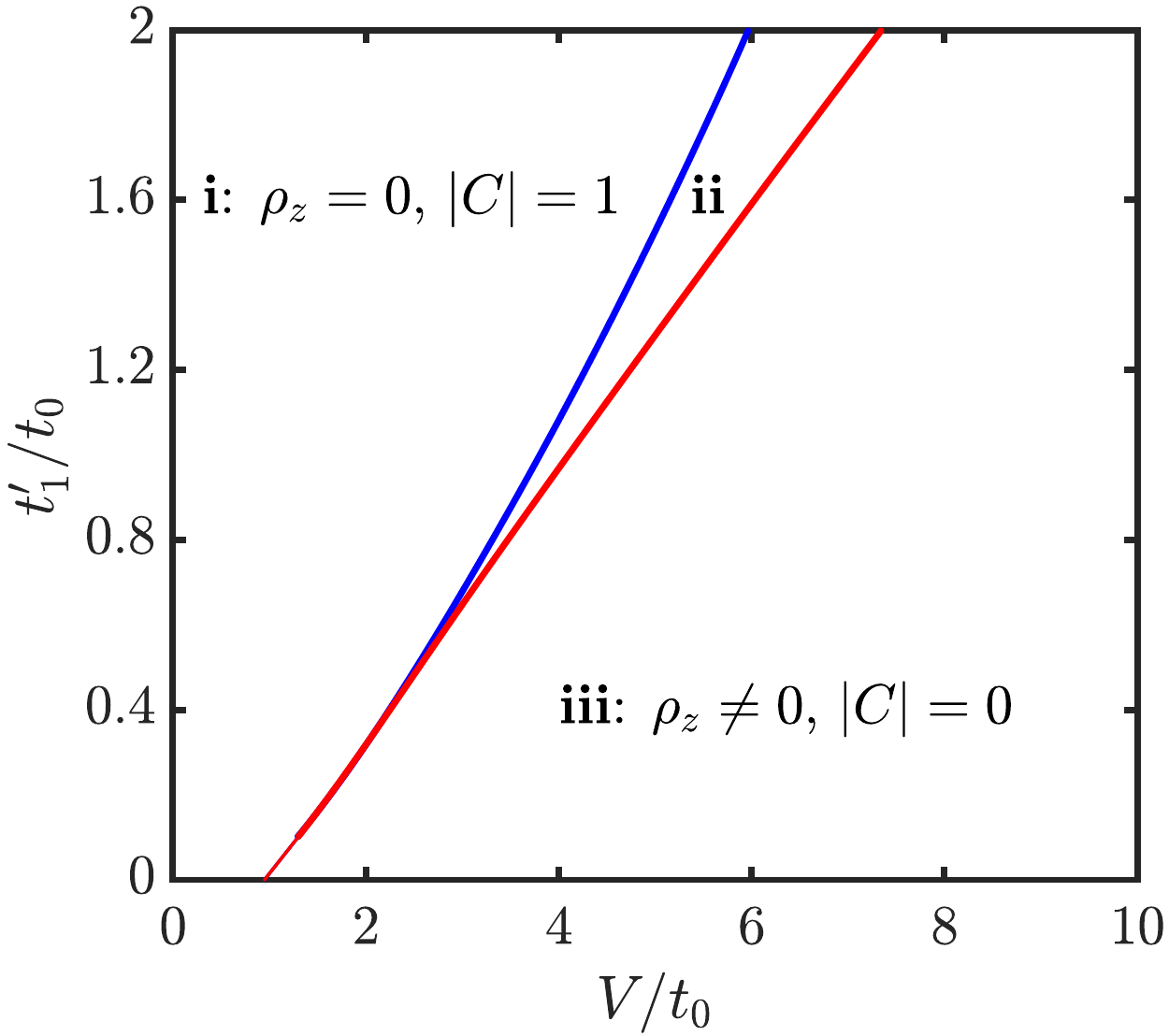}
\caption{Phase diagram of the Haldane-$V$ model. Phase $\bm{\textrm{i}}$ is the QAHI phase with $\rho_z=0$ and quantized Chern number $|C|=1$. Phase $\bm{\textrm{iii}}$ is sublattice polarized with  $\rho_z\neq 0$ and topologically trivial with $|C|=0$. There is an intermediate phase $\bm{\textrm{ii}}$ with $\rho_z\neq 0$ but $|C|=1$.}
\label{fig10}
\end{figure}

\begin{figure}[t]
\centering
\includegraphics[width=1.0\columnwidth,trim=3 0 0 0,clip]{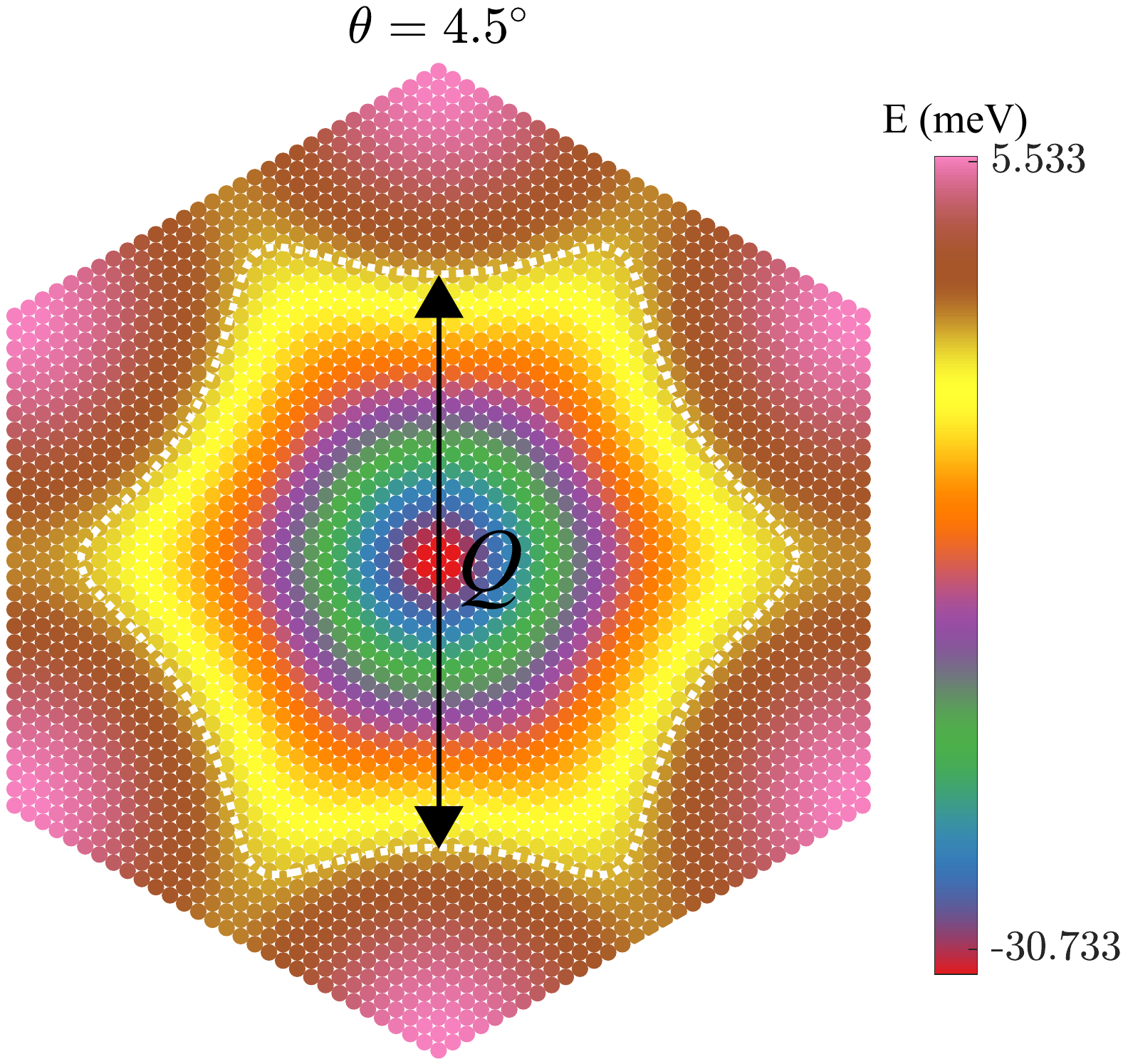}
\caption{Noninteracting Fermi surface (white dashed line) at half-filling of the topmost moiré valence bands. The color map shows the energy dispersion in the first moir\'e Brillouin zone. The vector $\mathcal{\bm{Q}}$ indicates an approximate nesting.}
\label{fig11}
\end{figure}

\subsection{Haldane-$V$ model}
The spontaneous sublattice (layer) polarization in the LP-FM$_z$ phase is driven primarily by the repulsion between $A$ and $B$ orbitals. To capture this physics, we focus on states with out-of-plane spin polarization. In this spin sector, we can study the spinless Haldane model with nearest-neighbor repulsion $V$, which is expressed as, 
\begin{equation}
\begin{aligned}
\hat{\mathcal{H}}&=\hat{\mathcal{H}}_\textrm{H}+\hat{\mathcal{H}}_V,\\
\hat{\mathcal{H}}_\textrm{H}&=\sum_{\langle i j\rangle} t_0 b_i^{+} b_j+\sum_{\langle\langle i j\rangle\rangle} t_{i j} b_i^{+} b_j \\
&=\sum_{\bm{k}} b_{\boldsymbol{k}}^{+}\left[F_0 \sigma_0+F_x \sigma_x+F_y \sigma_y+F_z \sigma_z\right] b_{\boldsymbol{k}}, \\
\hat{\mathcal{H}}_V&=V \sum_{\langle i j\rangle} \hat{n}_i \hat{n}_j ,
\end{aligned}
\end{equation}
where $\hat{\mathcal{H}}_\textrm{H}$ is the Haldane model on the honeycomb lattice and $\hat{\mathcal{H}}_V$ describes the nearest-neighbor repulsion. We note that $\hat{\mathcal{H}}_\textrm{KM}$ can be understood as two copies of $\hat{\mathcal{H}}_\textrm{H}$ that are time-reversal partners.

In the Hartree-Fock theory, $\hat{\mathcal{H}}$ can be approximated as,
\begin{equation}
\begin{aligned}
&\hat{\mathcal{H}} \approx 3 N V\left[\rho^2-\frac{\nu^2}{4}\right]+\sum_{\bm{k}}\left[F_0+\frac{3 V \nu}{2}\right]b_{\bm{k}}^{+}\sigma_0 b_{\bm{k}}\\
&+\sum_{\bm{k}} b_{\bm{k}}^{+}\left(\begin{array}{cc}
F_z-3 V \rho_z & \left(t_0-V \rho_{\|}\right) f_{\|} \\
\left(t_0-V \rho_{\|}^*\right) f_{\|}^* & -F_z+3 V \rho_z
\end{array}\right) b_{\boldsymbol{k}},
\end{aligned}
\end{equation}
where $b_{\boldsymbol{k}}=\left\{b_{\boldsymbol{k} A}, b_{\boldsymbol{k} B}\right\}^{\mathrm{T}}$ and $f_{\|}=F_{\|} / t_0$. We define a vector $\bm{\rho}$ as $\left\langle b^{+} \boldsymbol{\sigma} b\right\rangle/2$, where $b=\left\{ b_A,b_B\right\}^{\mathrm{T}}$ for a pair of nearest neighbors. $\rho_z$ is the $z$ component of $\boldsymbol{\rho}$ and quantifies the sublattice polarization. $\rho_{\|}$ is defined as $\rho_x-i\rho_y$ and quantifies the inter-sublattice coherence. The nearest-neighbor hopping parameter is renormalized from the bare value $t_0$ to be $t_0-V\rho_{\|}$ by the interaction.

\begin{figure}[t]
\centering
\includegraphics[width=1.0\columnwidth,trim=3 0 0 0,clip]{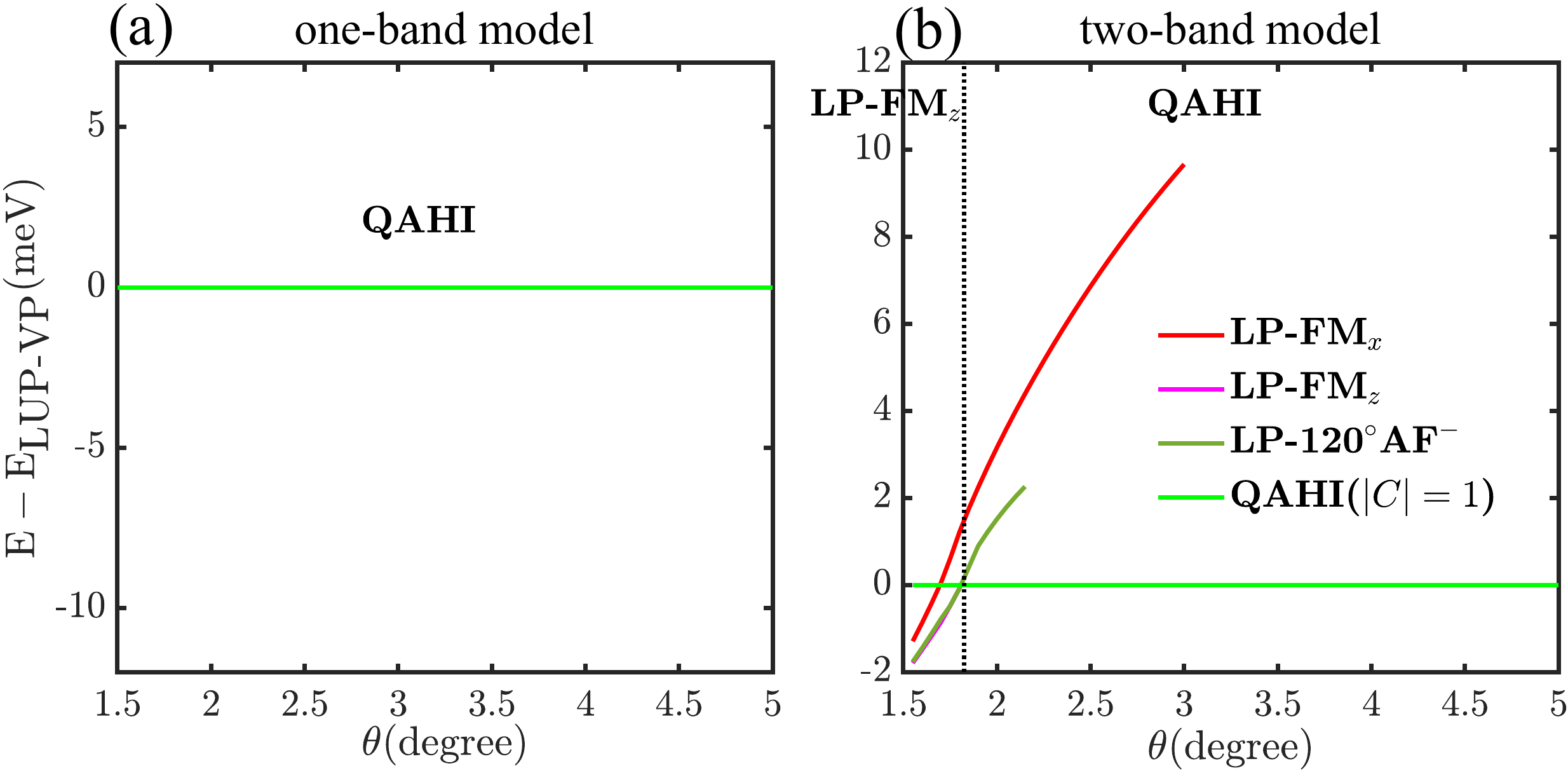}
\caption{Energy per moir\'e unit cell of competing states relative to the LUP-VP states at $\nu=1$, $\epsilon = 15$ and $d = 20$ nm. The results are obtained using (a) one-band and (b) two-band models, respectively. Other parameter values are the same as those used in Fig.~\ref{fig3}(b). }
\label{fig8}
\end{figure}

The total mean-field energy at $\nu=1$ is,
\begin{equation}
\begin{aligned}
E_{tot}\left(\rho_z, \rho_{\|}\right)&=\frac{3}{4} V N+\sum_{\bm{k}} F_0+3 V N \rho^2 \\
&-\sum_{\bm{k}} \sqrt{\left(F_z-3 V \rho_z\right)^2+\left|\left(t_0-V \rho_{\|}\right) f_{\|}\right|^2},
\end{aligned}
\end{equation}
where $\rho^2=\left|\rho_{\|}\right|^2+\rho_z^2$. We obtain the following two self-consistent equations by, respectfully, taking $\frac{\partial E_{tot}}{\partial \rho_z}=0$ and $\frac{\partial E_{tot}}{\partial \rho_{\|}}=0$,
\begin{equation}
\begin{aligned}
\rho_z=\frac{1}{2 N} \sum_{\bm{k}} \frac{3 V \rho_z-F_z}{\sqrt{\left(F_z-3 V \rho_z\right)^2+\left|\left(t_0-V \rho_{\|}\right) f_{\|}\right|^2}},\\ 
\rho_{\|}=\frac{1}{6 N} \sum_{\bm{k}} \frac{\left(V \rho_{\|}-t_0\right)\left|f_{\|}\right|^2}{\sqrt{\left(F_z-3 V \rho_z\right)^2+\left|\left(t_0-V \rho_{\|}\right) f_{\|}\right|^2}}.
\end{aligned}
\end{equation}

We note that the self-consistent equations always have a symmetric solution with no sublattice polarization ($\rho_z=0$), but it can also have a symmetry-breaking solution with spontaneous sublattice polarization ($\rho_z\neq 0$). By minimizing the total energy, we obtain the phase diagram [Fig.~\ref{fig10}] of the Haldane-$V$ model as a function of $V/t_0$ and $t'_1/t_0$, where $t'_1$ is the imaginary part of one of the second nearest neighbor hopping parameters. For a fixed $t'_1/t_0$, $\rho_z$ is 0 (nonzero) for $V$ below (above) a critical value. Therefore, spontaneous sublattice polarization can be driven by $V$ above a critical value. This qualitatively explains why the LP-FM$_z$ becomes the ground state at small twist angles, as $V/t_0$ increases with decreasing twist angle. The state with a spontaneous finite  $\rho_z$ can also be described as a charge density wave state with imbalanced charge density at $A$ and $B$ sites. In $t$MoTe$_2$, $A$ and $B$ orbitals have opposite layer polarization, and therefore, spontaneous sublattice polarization also leads to spontaneous layer polarization.

\subsection{Fermi surface instability}\label{fsi}
At large twist angles, the Coulomb interaction can become less dominant compared to the single-particle kinetic energy, and the fermiology associated with the Fermi surface can become important. In Fig.~\ref{fig11}, we plot the noninteracting Fermi surface at half-filling (i.e., $\nu=1$) of the topmost moir\'e valence bands for $\theta=4.5^\circ$. There is an approximate Fermi surface nesting, and the nesting vector $\mathcal{\bm{Q}}$ is approximately equal to $\bm\kappa_{-}-\bm\kappa_{+}$, which can drive Fermi surface instability with a $\sqrt{3}\times\sqrt{3}$ real-space texture. This instability exactly explains the formation of the $O-120^{\circ}$AF$^{\pm}$ state at large twist angles.

\begin{figure}[t]
\centering
\includegraphics[width=1.0\columnwidth,trim=3 0 0 0,clip]{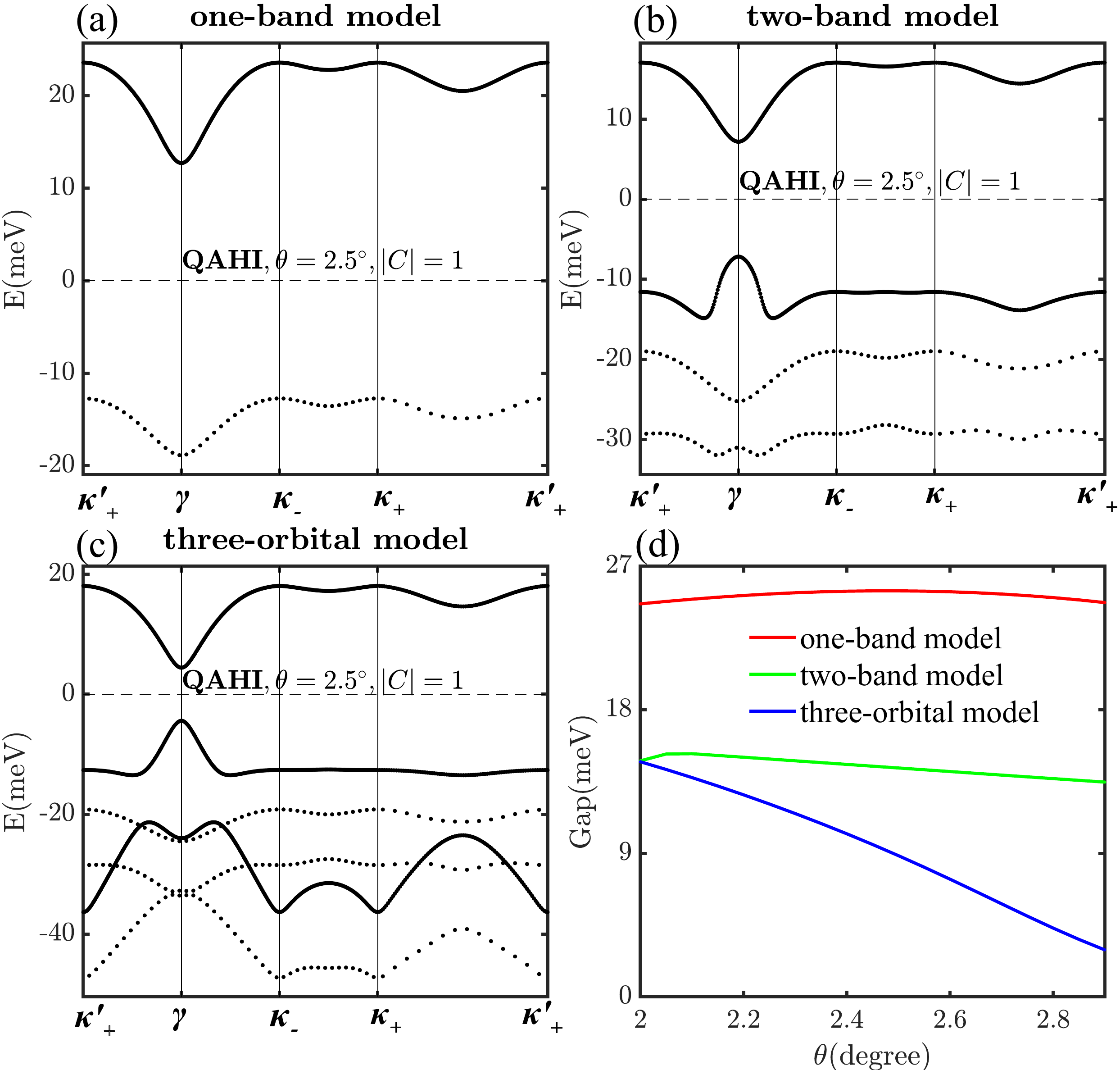}
\caption{(a)-(c) The HF band structure for the $\nu=1$ QAHI phase at $\theta = 2.5^\circ$ calculated with one-band, two-band, and three-orbital model, respectively. The solid and dotted lines, respectively, plot bands from $+K$ and $-K$ valleys. (d) The charge gap of the QAHI phase obtained from one-band (red), two-band (green) and three-orbital (blue) model for the twist angle range from $2^\circ$ to $2.9^\circ$, where the QAHI phase is the same ground state in all three models.  Parameter values are the same as those used in Fig.~\ref{fig8}.}
\label{fig9}
\end{figure}

\section{Comparison of different models}\label{appc}
In this Appendix, we present a comparison of results obtained, respectively, from one-band, two-band, and three-orbital models, as summarized in Fig.~\ref{fig8}. First, in the one-band model [Fig.~\ref{fig8}(a)], we project the interacting Hamiltonian onto the topmost moir\'e valence bands, where each valley contributes one band. As shown in Fig.~\ref{fig8}(a), the valley polarized state at $\nu=1$ is always the QAHI in the one-band model. This model fails to capture the LP-FM$_z$ state, the topologically trivial FM$_z$ state, and the $O-120^{\circ}$AF$^{\pm}$ state. (2) In the two-band model, we project the interacting Hamiltonian onto the first two moir\'e valence bands, where each valley contributes two bands. As shown in Fig.~\ref{fig8}(b), the two-band model captures the LP-FM$_z$ state for small twist angles and its transition to the QAHI state above a critical angle, but fails to describe the topologically trivial FM$_z$ and $O-120^{\circ}$AF$^{\pm}$ states. Quantitatively, the transition angle between the LP-FM$_z$ and QAHI states is underestimated in the two-band model compared to the three-orbital model shown in Fig.~\ref{fig3}(b). The additional $O$ orbital in the three-orbital model is crucial in correctly finding all competing states such the topologically trivial FM$_z$ and the $O-120^{\circ}$AF$^{\pm}$ states.

The QAHI state is the same ground state for one-band, two-band model, and three-orbital models for a certain range of twist angles. However, there is a crucial difference in the size of the charge gap. As shown in Fig.~\ref{fig9}, the one-band and two-band models generally overestimate the charge gap compared to the three-orbital model. In the three-orbital model, the charge gap has a charge-transfer nature, since it is determined by electron and hole located at different real-space orbitals but within the same valley. The one-band model completely fails to capture this charge-transfer gap, and therefore, overestimates the charge gap.

We also perform the Hartree-Fock calculation in the plane-wave basis without projecting onto the first three bands, which leads to quantum phase diagrams that quantitatively agree with that from the three-orbital model, as we report in a follow-up work \cite{li2023electrically}. Therefore, the three-orbital model can be viewed as a minimal model that provides a unified description of competing phases over a large range of twist angles.

\bibliography{reference}

\end{document}